\documentclass[11pt]{article}
\usepackage{amsmath,amssymb,amsthm,mathrsfs,csquotes,physics,bm}
\usepackage{color}

\usepackage{graphicx}
\graphicspath{{figure/}}

\usepackage{float}
\usepackage{tikz}
\usepackage{verbatim}
\usepackage[normalem]{ulem}

\numberwithin{equation}{section}

\newcommand{\pd}{\partial}
\newcommand{\TTbar}{T\overline{T}}

\usepackage[numbers,sort&compress]{natbib}
\usepackage[colorlinks=true,      
linkcolor=blue,       
citecolor=green,      
urlcolor=cyan,        
bookmarks=true,       
bookmarksopen=true]{hyperref}  

\begin{document}

	\setlength{\hoffset}{-1in} 
	\setlength{\oddsidemargin}{.14\paperwidth}          
	\setlength{\evensidemargin}{.14\paperwidth}         
	\setlength{\marginparwidth}{.11\paperwidth}         
	\setlength{\textwidth}{.72\paperwidth}              
	\setlength{\voffset}{-1in}  
	\setlength{\topmargin}{.05\paperheight}            
	\setlength{\headheight}{.02\paperheight}           
	\setlength{\headsep}{.03\paperheight}                
	\setlength{\textheight}{.76\paperheight}              
	\setlength{\footskip}{.07\paperheight}                
	\setlength{\parskip}{0pt}                           
	

\thispagestyle{empty}

\begin{center}
		{\Large\textbf{\mathversion{bold}
				Stress Tensor Deformations in dS/CFT: Mixed Boundary Conditions, Spectrum Flow and Pseudo Entropy 
			}
			\par}
		
		\vspace{0.5cm}
		
		{ Feng Hao$^{1,*}$, Hao Ouyang$^{1,*}$ and Xi-Yang Ran$^{1,*}$ 
		}\\
		\vspace*{0.2cm}
		{\it
			$^{1}$Center for Theoretical Physics and College of Physics, Jilin University, 
			Changchun 130012, China\\
		}
		
		\vspace*{0.5cm}
		{E-mails: {\tt haofeng22@mails.jlu.edu.cn, haoouyang@jlu.edu.cn,  ranxy23@mails.jlu.edu.cn. 
		}}
		\vspace{1cm}

		

		\par\vspace{1.5cm}
		
\textbf{Abstract} \vspace{3mm}
		
\begin{minipage}{\textwidth}
			
We formulate a local semiclassical source--response prescription for derivative-free stress-tensor deformations in dS/CFT. The metric flow determines the deformed boundary metric and stress tensor, encoding the deformation through mixed boundary conditions at future infinity. In Kerr-dS$_3$/CFT$_2$, the holographic charges reproduce the field-theory spectral flow, providing an internal consistency check. For general deformations in this class, we derive a holographic pseudo-entropy formula in terms of the flowed boundary data, nonperturbative in the deformation parameters within a specified complex-geodesic prescription. We evaluate this formula explicitly for the $T\bar{T}$ and root-$T\bar{T}$ deformations. On the non-rotating principal branch, an analytically continued replica calculation reproduces the interval-dependent corrections at first order in the couplings and leading semiclassical order. Agreement of the interval-independent terms additionally requires a source-frame regulator identification.

\vspace{10pt}

\end{minipage}
		
\end{center}
	
	
	\vspace{1.5cm}
	
	\begingroup
	\renewcommand\thefootnote{*}
	\footnotetext{Corresponding authors.}
	\addtocounter{footnote}{0}
	\endgroup

	\newpage
	{
		\hypersetup{linkcolor=black}
		\tableofcontents
	}
	\newpage	
	
	
	\section{Introduction}
	
	The dS/CFT correspondence proposes that quantum gravity in asymptotically de Sitter spacetime is encoded by a Euclidean conformal field theory at future infinity \cite{Strominger:2001pn,Witten:2001kn,Balasubramanian:2001nb,Klemm:2001ea,Balasubramanian:2002zh,Anninos:2012qw}.
	This setting is subtler than AdS/CFT: the asymptotic boundary is spacelike, the holographic direction is tied to cosmological time, and the natural bulk object is the wavefunction of the universe rather than a Lorentzian boundary partition function \cite{Maldacena:2002vr,Harlow:2011ke}.
	Boundary data and boundary conditions are therefore integral to the holographic dictionary.
	A geometric realization of boundary deformations must, in particular, specify how they modify the asymptotic data at future infinity.
	
	A useful lesson from AdS holography is that deformations by stress-tensor operators can often be formulated by replacing Dirichlet boundary conditions for the bulk metric with mixed ones.
	In two dimensions, the $\TTbar$ deformation is a solvable irrelevant deformation generated by a composite stress-tensor operator \cite{Zamolodchikov:2004ce,Smirnov:2016lqw,Cavaglia:2016oda} (see \cite{Jiang:2019epa,He:2025ppz} for reviews).
	Its holographic descriptions include finite-cutoff AdS$_3$ \cite{McGough:2016lol,Kraus:2018xrn} and, more generally, mixed boundary conditions for the bulk metric \cite{Guica:2019nzm}.
	Related stress-tensor deformations and higher-dimensional generalizations have been studied in
	\cite{Hartman:2018tkw,Taylor:2018xcy,Bonelli:2018kik,Babaei-Aghbolagh:2022leo,Ferko:2022cix,Babaei-Aghbolagh:2020kjg,Babaei-Aghbolagh:2024hti,Ferko:2023sps,Ferko:2024yua}.
	A particularly useful formulation is the metric-flow approach, which evolves the background metric and stress tensor
	\cite{Conti:2022egv,Morone:2024ffm,Ran:2024vgl,Morone:2024sdg}. 
	Holographically, these flows can be interpreted as mixed boundary conditions for the bulk metric
	\cite{Guica:2019nzm,Ebert:2023tih,Ran:2025xas}. 
	The deformation is thus encoded in the relation between the boundary metric and stress tensor.
	
This boundary-condition viewpoint is especially useful in de Sitter holography: the data at future infinity remain natural, whereas a literal finite-radius cutoff is less direct. 
Nevertheless, de Sitter cutoff and moving-boundary constructions include Cauchy-slice holography and its dS realization \cite{Araujo-Regado:2022cqf,Araujo-Regado:2024xxx}, timelike-screen holography \cite{Kawamoto:2023nki}, and the curved-space Wheeler--DeWitt formulation of the $\TTbar$ flow \cite{Mazenc:2019zzz}. 
The $T^2$ deformation of \cite{Araujo-Regado:2024xxx} provides a complementary construction on flat Cauchy slices with emergent time. 
Other perspectives on stress-tensor deformations and $\TTbar+\Lambda_2$-type flows \cite{Lewkowycz:2019xse,Gorbenko:2018oov} in de Sitter holography appear in \cite{Shyam:2021ciy,Coleman:2021nor,Torroba:2022jrk,Batra:2024kjl,Chen:2023eic,Aguilar-Gutierrez:2024nst,Chang:2024voo,Chang:2025ays}. 
These developments motivate our central questions: can stress-tensor deformations in dS/CFT be formulated systematically as mixed boundary conditions at future infinity, and how does the bulk realize the resulting boundary data? 
	
	These features also change the interpretation of holographic entropy in de Sitter space.
	Because the natural dS/CFT boundary object is a cosmological wavefunction, cutting and gluing its path integral need not produce Hermitian density matrices \cite{Strominger:2001pn,Witten:2001kn,Maldacena:2002vr,Harlow:2011ke}.
	This is reflected in complex geodesic lengths and complex-valued holographic entropies \cite{Narayan:2015vda,Narayan:2015oka,Doi:2022iyj,Narayan:2022afv}, now more naturally interpreted as pseudo entropies of reduced transition matrices than as ordinary entanglement entropies \cite{Nakata:2020luh,Mollabashi:2020yie,Mollabashi:2021xsd,Doi:2022iyj,Doi:2023zaf,Narayan:2022afv,Mukherjee:2022jac,Zhao:2025zgm}.
	Complex extremal surfaces and geodesics thereby probe transition-amplitude-like data in the putative dual theory.
	
	Three-dimensional de Sitter gravity provides a controlled setting for these questions, and the dS$_3$/CFT$_2$ dictionary has recently been sharpened from several perspectives \cite{Balasubramanian:2001nb,Hikida:2022ltr}.
	Although pure three-dimensional Einstein gravity has no local propagating degrees of freedom, it admits nontrivial quotients characterized by global charges.
	Kerr-dS$_3$ is a simple example: it is locally de Sitter, contains a cosmological horizon, and carries mass and angular momentum through a spinning conical defect \cite{Park:1998qk,Balasubramanian:2001nb}.
	In the putative dual theory, these global parameters map to state energy, momentum, and degeneracy.
	Kerr-dS$_3$ therefore tests the proposed stress-tensor deformation dictionary through both the boundary spectrum and holographic pseudo entropy.
	
In this work, we formulate stress-tensor deformations in dS/CFT through mixed boundary conditions at future infinity, using the semiclassical metric-flow framework. Applying this construction to Kerr-dS$_3$, we determine the deformed geometry and holographic charges and show that the latter agree with the field-theory spectral flow. We also obtain finite-coupling expressions for holographic pseudo entropy under the $\TTbar$ and root-$\TTbar$ deformations and analyze competing connected geodesics in the non-rotating sector. An analytically continued boundary replica calculation reproduces the interval-dependent corrections at first order in both couplings and leading semiclassical order.

Sec.~\ref{sec:flow eq} develops the metric-flow and mixed-boundary-condition framework. Sec.~\ref{sec:deformed dS3} applies it to Kerr-dS$_3$/CFT$_2$ and compares the holographic charges with the field-theory spectrum. Sec.~\ref{sec:EE} studies holographic pseudo entropy, and Sec.~\ref{sec:Conclusion} presents our conclusions and outlook.
App.~\ref{app:complex-geodesic-nonrotating} and \ref{app:replica pseudo entropy} present the detailed geodesic analysis and the perturbative boundary replica calculation, respectively.

	
	\section{Metric flow and mixed boundary conditions in dS/CFT}
	\label{sec:flow eq}
	
	\subsection{Stress tensor deformations and metric flow}
	\label{subsec:metric-flow}
	
	We first review the metric-flow formulation of stress-tensor deformations, following \cite{Conti:2022egv,Ran:2025xas}.
	Consider a $d$-dimensional Euclidean field theory on a background metric $\gamma_{ab}$, deformed by a scalar operator $\mathcal{O}=\mathcal{O}(T^{a}{}_b;\lambda)$ constructed from the stress tensor $T^{ab}$.
	The action obeys
	\begin{equation}\label{eq:def of deformation diff eq}
		\frac{\partial S_\lambda[\gamma,\phi]}{\partial\lambda}
		=
		\int \dd^d x\,\sqrt{\gamma} \mathcal O~,
		\qquad
		\gamma\equiv |\det\gamma_{ab}|~,
	\end{equation}
	where $\phi$ collectively denotes the matter fields and $\lambda$ is the deformation parameter.
	We take the deformation operator to be a scalar function of the mixed tensor
	\begin{equation}
		T^a{}_b\equiv T^{ac} \gamma_{bc}~,
	\end{equation}
	constructed from the stress tensor defined by 
	\begin{equation}\label{eq:def of stress tensor}
		T^{ab}=-\frac{2}{\sqrt{\gamma}}\frac{\delta S_\lambda[\gamma,\phi]}{\delta\gamma_{ab}}~.
	\end{equation}
	
	More precisely, throughout the general discussion we restrict to
	derivative-free, similarity-invariant scalar functions of the mixed
	stress tensor,
	\begin{equation}
		\mathcal O
		=
		\mathcal F(I_1,\ldots,I_d;\lambda),
		\qquad
		I_k\equiv
		\tr\!\left[(T^a{}_b)^k\right].
		\label{eq:class-function-O}
	\end{equation}
	We assume that $\mathcal F$ is locally differentiable or defined on a specified analytic branch.
	Deformations involving stress-tensor derivatives, curvature tensors, or additional background sources require an enlarged characteristic system and lie outside the eigenvalue reduction below.

At large $N$, the saddle-point action dominates the generating functional, and factorization replaces composite stress-tensor operators by functions of one-point data.
The resulting first-order Hamilton--Jacobi equation yields, by the method of characteristics, coupled flows for the source $\gamma_{ab}$ and the stress-tensor one-point function \cite{Guica:2019nzm,Conti:2022egv,Ebert:2023tih,Morone:2024ffm,Ran:2024vgl,Morone:2024sdg}:
	\begin{align}
		\frac{\partial \gamma_{ab}}{\partial\lambda}&=2\,\frac{\partial \mathcal O}{\partial T^{ab}}~,\label{eq:flow eq for gamma}\\
		\frac{\partial T^{ab}}{\partial\lambda}&=\left(T^{cd}\gamma^{ab}-T^{ab}\gamma^{cd}\right)\frac{\partial\mathcal O}{\partial T^{cd}}-2\,\frac{\partial\mathcal O}{\partial\gamma_{ab}}-\mathcal O\,\gamma^{ab}~.\label{eq:flow eq for stress tensor}
	\end{align}
	Here, the derivative
	$\partial\mathcal O/\partial T^{ab}$ is taken while fixing $\gamma_{ab}$ and vice versa. 
	These equations describe the source $\gamma_{ab}$ and one-point-function $T^{ab}$ flows induced by \eqref{eq:def of deformation diff eq}. They do not yet invoke bulk dynamics.
	
	Equivalently, the same equations may be obtained by requiring that the deformation preserve the canonical variational form
	\begin{equation}
		\delta S_\lambda=-\frac12\int d^dx\sqrt\gamma\,T^{ab}\delta\gamma_{ab},
	\end{equation}
	with the variation commuting with the deformation flow. This variational derivation is the standard source--response derivation used in the mixed-boundary-condition literature \cite{Guica:2019nzm,Ebert:2023tih,Ran:2025xas}. In the present work we use the characteristic derivation summarized in App.~\ref{app:character}.
	
	Given undeformed data $\gamma_{ab}^{[0]},T^{ab}_{[0]}$, Eqs.~\eqref{eq:flow eq for gamma} and \eqref{eq:flow eq for stress tensor} determine the deformed metric and stress tensor.
	Holographically, these data specify mixed boundary conditions for the bulk metric.
	
	\subsection{Mixed boundary conditions from metric flow}
	\label{subsec:mixed-bc}
	
	We now give the holographic interpretation of the metric-flow equations.
	The relation between multi-trace deformations and modified boundary conditions is standard in AdS/CFT
	\cite{Klebanov:1999tb,Witten:2001ua,Berkooz:2002ug}. 
	Systematic treatments of the large-$N$ relation between sources, one-point functions, and the generating functional appear in \cite{Mueck:2002gm,Gubser:2002vv,Hartman:2006dy,Diaz:2007an,Papadimitriou:2007sj}.
	With a semiclassical gravitational dual, the boundary metric sources the stress tensor, whose one-point function is the renormalized Brown--York tensor.
	A stress-tensor deformation changes this source--response relation and therefore the boundary conditions for the bulk metric.
	
	At large $N$, let $\gamma^{[0]}_{ab}$ and $T^{ab}_{[0]}$ denote the metric and stress tensor of the undeformed theory.
	In the standard Dirichlet situation, $\gamma^{[0]}_{ab}$ is fixed as the boundary source, while $T^{ab}_{[0]}$ is determined by the normalizable component of the bulk solution. 
	After turning on the deformation, Eqs.~\eqref{eq:flow eq for gamma} and \eqref{eq:flow eq for stress tensor} determine a new pair of boundary data,
	\begin{equation}
		\gamma^{[\lambda]}_{ab}=\gamma_{ab}~, \quad T^{ab}_{[\lambda]}=T^{ab}~,
	\end{equation}
	with initial condition
	\begin{equation}
		\gamma_{ab}|_{\lambda=0}=\gamma^{[0]}_{ab}~, \quad T^{ab}|_{\lambda=0}=T^{ab}_{[0]}~.
	\end{equation}
	The holographic prescription imposes at the asymptotic boundary the relation between $\gamma^{[\lambda]}_{ab}$ and $T^{ab}_{[\lambda]}$ generated by the flow.
	This condition is mixed because it fixes a specific combination of the boundary metric and Brown--York tensor, rather than either quantity separately.

This formulation differs from a finite-cutoff picture.
	For the ordinary two-dimensional $\TTbar$ deformation on the boundary of pure AdS$_3$ gravity, the mixed boundary condition can, for one sign of the deformation parameter, be rewritten on shell as a Dirichlet condition at a finite radial cutoff surface \cite{Guica:2019nzm}.
	This equivalence relies only on the truncated three-dimensional Fefferman--Graham expansion and therefore extends to constant-curvature saddles of either sign, including pure dS$_3$ and the Kerr-dS$_3$ family of Sec.~\ref{sec:deformed dS3}.
	Finite-cutoff realizations also exist in higher dimensions \cite{Hartman:2018tkw}, but mixed boundary conditions are more general.
	For $\TTbar$, the cutoff interpretation holds only for one sign of the coupling (Footnote~\ref{fn:cutoff}). Generic stress-tensor deformations admit no such interpretation \cite{Ebert:2023tih,Ran:2025xas}.
	Bulk matter also breaks the equivalence: a radial cutoff imposes Dirichlet conditions on all bulk fields, whereas the deformation changes only the metric boundary condition \cite{Guica:2019nzm}.
	This distinction will be important below, since in de Sitter holography the natural boundary data live at future infinity rather than at a spatial radial cutoff surface.

Solving the metric-flow equations therefore determines the asymptotic source--response relation. 
A bulk solution satisfying it is interpreted as dual to the stress-tensor-deformed boundary theory.
	
\subsection{Analytic continuation to dS/CFT}\label{subsec:ads-to-ds}
	
We now adapt the mixed-boundary-condition prescription to dS/CFT.
	In AdS/CFT, the Gubser--Klebanov--Polyakov--Witten (GKPW) relation identifies the renormalized on-shell action with the boundary generating functional \cite{Gubser:1998bc,Witten:1998qj}. At large $N$,
	\begin{equation}\label{eq:AdS GKPW}
		Z_{\rm AdS}[\gamma_{ab}] \simeq \exp\left(-S^{\rm AdS}_{\rm ren}[\gamma_{ab}] \right)~.
	\end{equation}
	Here $S^{\rm AdS}_{\rm ren}[\gamma_{ab}]$ is the renormalized Euclidean bulk on-shell action with boundary source $\gamma_{ab}$.
	
	In dS/CFT, the corresponding semiclassical object is instead the cosmological wavefunction with prescribed data at future infinity \cite{Strominger:2001pn,Witten:2001kn,Maldacena:2002vr,Harlow:2011ke},
	\begin{equation}
		\Psi_{\rm dS}[\gamma_{ab}]
		\simeq
		\exp\left(i S_{\rm ren}^{\rm dS}[\gamma_{ab}]\right)~.
		\label{eq:dS wavefunction}
	\end{equation}
	Here $S_{\rm ren}^{\rm dS}$ is the renormalized on-shell action evaluated on a chosen bulk saddle and analytic branch, with boundary metric $\gamma_{ab}$ at $\mathcal I^+$.
	Equation~\eqref{eq:dS wavefunction} fixes our phase convention for this saddle-level prescription.
	
The real Brown--York response entering the mixed boundary condition should be distinguished from the stress-tensor one-point function defined by the dS wavefunction. 
We use
\begin{align}
    T_{\rm BY}^{ab}
    &\equiv
    -\frac{2}{\sqrt\gamma}
    \frac{\delta S_{\rm ren}^{\rm dS}}{\delta\gamma_{ab}}~,
    \\
    I_{\Psi}[\gamma]
    &\equiv
    -\log\Psi_{\rm dS}[\gamma]
    =-iS_{\rm ren}^{\rm dS}[\gamma]~,
    \\
    T_{\Psi}^{ab}
    &\equiv
    -\frac{2}{\sqrt\gamma}
    \frac{\delta I_{\Psi}}{\delta\gamma_{ab}}
    =
    \frac{2}{\sqrt\gamma}
    \frac{\delta\log\Psi_{\rm dS}}{\delta\gamma_{ab}}
    =-iT_{\rm BY}^{ab}~.
\end{align}
Thus $T_{\rm BY}^{ab}$ is the real gravitational response used in the mixed-boundary-condition flow, whereas $T_{\Psi}^{ab}$ is the boundary one-point function in the wavefunction convention. 
Below, an unsubscripted $T^{ab}$ in the bulk construction denotes $T_{\rm BY}^{ab}$.

The semiclassical dS prescription is related to the AdS one by analytic continuation. 
For the curvature radius,
\begin{equation}\label{eq:AdS to dS continuation}
    \ell_{\rm AdS}^{2}
    \longrightarrow
    -\ell_{\rm dS}^{2}~,
    \qquad
    \ell_{\rm AdS}
    \longrightarrow
    -i\ell_{\rm dS}~,
\end{equation}
where the second relation specifies the branch used below, together with the corresponding continuation of the radial coordinate. 
We denote this saddle-level continuation by $\mathcal C$. 
On this branch,
\begin{align}
    Z_{\rm AdS}[\gamma]
    &\xrightarrow{~\mathcal C~}
    \Psi_{\rm dS}[\gamma]~,
    \\
    S_{\rm ren}^{\rm AdS}
    &\xrightarrow{~\mathcal C~}
    I_{\Psi}
    =-iS_{\rm ren}^{\rm dS}~,
    \\
    T_{\rm AdS}^{ab}
    \equiv
    \frac{2}{\sqrt\gamma}
    \frac{\delta\log Z_{\rm AdS}}{\delta\gamma_{ab}}
    &\xrightarrow{~\mathcal C~}
    T_{\Psi}^{ab}
    =-iT_{\rm BY}^{ab}~,
    \\
    c_{\rm AdS}
    =\frac{3\ell_{\rm AdS}}{2G}
    &\xrightarrow{~\mathcal C~}
    c_{\Psi}
    =-ic_{\rm dS}~,
    \qquad
    c_{\rm dS}\equiv\frac{3\ell_{\rm dS}}{2G}>0~.
\end{align}
Here $c_{\rm dS}$ is the real dS central charge in the Brown--York convention, while $c_{\Psi}=-ic_{\rm dS}$ is its wavefunction counterpart. 
The continuation $\mathcal C$ refers only to the chosen classical saddle and does not specify the full gravitational integration contour.

In the holographic construction below, $\lambda$ denotes the real parameter of the Brown--York/mixed-boundary-condition flow. 
For a deformation operator $\mathcal O$, the classical flow is
\begin{equation}
    \pd_{\lambda}S_{\rm ren}^{\rm dS}
    =
    \int\dd^d x~\sqrt\gamma
    \mathcal O_{\rm BY}[T_{\rm BY}]~.
\end{equation}
Using $I_{\Psi}=-iS_{\rm ren}^{\rm dS}$ and $T_{\rm BY}=iT_{\Psi}$, the same flow can be written as
\begin{equation}
    \pd_{\lambda}I_{\Psi}
    =
    -i\int\dd^d x~\sqrt\gamma\,
    \mathcal O_{\rm BY}[iT_{\Psi}]~.
\end{equation}
The coupling $\mu$ multiplying the conventionally normalized operator in the wavefunction functional is therefore related to $\lambda$ in an operator-dependent way. 
Polynomial operators acquire the corresponding powers of $i$, while non-analytic operators additionally require a branch choice. 
For the $\TTbar$ and root-$\TTbar$ deformations, the explicit relations are given in Sec.~\ref{sec:TT entropy}. 
Below we write $\ell\equiv\ell_{\rm dS}$ unless stated otherwise. 

In pure three-dimensional Einstein gravity, the local metric-flow solutions and the deformed saddle admit continuation from AdS$_3$ on the branch \eqref{eq:AdS to dS continuation}, with the corresponding continuation of boundary time and quotient parameters. 
The square-root spectrum, the interval-dependent geodesic expression, and the discriminant defining the Hagedorn-like endpoint likewise inherit their functional forms from the
AdS$_3$ construction. 
Their de Sitter realization additionally requires a real Kerr-dS$_3$ section and compact quotient, a charge convention at the spacelike boundary, a fixed-state prescription, and a choice of complex-geodesic image and sheet. 
These inputs are specified in Secs.~\ref{sec:deformed dS3} and \ref{sec:EE}.

The boundary gravitational phase space provides a useful language for these deformations. 
In the semiclassical wavefunction, the boundary metric $\gamma_{ab}$ and its conjugate momentum $P^{ab}$ form a phase-space pair, with $P^{ab}$ proportional to the stress-tensor one-point function. 
A stress-tensor deformation changes their source--response relation and may therefore be viewed semiclassically as a canonical-transformation-like map,
\begin{equation}
	(\gamma_{ab},P^{ab})\longrightarrow(\gamma^{[\lambda]}_{ab},P^{ab}_{[\lambda]})~.
	\end{equation}
Related canonical perspectives on stress-tensor deformations, especially the $\TTbar$ case, have appeared in \cite{Jorjadze:2020ili,Benitez:2023ojg}. 
	
At the quantum level, the deformation should act directly on the cosmological wavefunction. 
With a wavefunction-convention coupling $\mu$, one may write schematically
\begin{equation}
	\partial_{\mu}\Psi_{\mu}[\gamma]=\widehat{\mathcal D}[\gamma,\widehat P]~\Psi_{\mu}[\gamma]~,\qquad\widehat P^{ab}\sim-i\frac{\delta}{\delta\gamma_{ab}}~.
\end{equation}
Here $\widehat{\mathcal D}$ is fixed by the deformation, with the replacement of the stress tensor by the momentum operator understood schematically. 
The operator-dependent map $\mu(\lambda)$ may be complex and is derived for our two deformations at the beginning of Sec.~\ref{sec:TT entropy}.
A complete definition would also specify operator ordering, contact terms, the functional measure, and compatibility with gravitational constraints.
Equivalently, an integral kernel can represent the change from $\gamma$ to the deformed source $\Gamma\equiv\gamma^{[\lambda]}(\gamma,P)$ associated with the real mixed-boundary-condition parameter:
\begin{equation}\label{eq:formal wavefunction kernel}
	\Psi_{\mu(\lambda)}[\Gamma]=\int D\gamma~K_\lambda[\Gamma,\gamma]\Psi_0[\gamma]~.
\end{equation}
In the semiclassical limit,
\begin{equation}
	K_\lambda[\Gamma,\gamma]\sim\exp\left[\frac{1}{G}\mathcal W_\lambda[\Gamma,\gamma]\right]~,
\end{equation}
where $\mathcal{W}_\lambda$ is generally complex in dS/CFT. 
The saddle-point equations
\begin{equation}
	P^{ab}=\frac{i}{G}\frac{\delta \mathcal W_\lambda}{\delta\gamma_{ab}}~,\qquad \Pi^{ab} = -\frac{i}{G}\frac{\delta \mathcal W_\lambda}{\delta\Gamma_{ab}},
\end{equation}
relate sources and one-point functions classically.
The metric-flow source--response relation just corresponds to this semiclassical limit.
Even for real Brown--York/mixed-boundary-condition parameters, the wavefunction couplings need not be real. 
In either parametrization, the kernel generally implements a change of boundary condition or wavefunction representation rather than unitary evolution.
We leave the full quantum kernel to future work and use its classical source--response map to prescribe mixed boundary conditions at $\mathcal I^+$.
	
This phase-space language also helps compare de Sitter holographic setups.
Future-boundary dS/CFT fixes data on a spacelike surface near $\mathcal I^+$, whereas static-patch and stretched-horizon descriptions involve timelike screens or a static observer's worldline \cite{Anninos:2011af,Nakayama:2011qh}. 
Reference~\cite{Chang:2025ays} associates inward motion of a spacelike boundary with a $\TTbar$ flow and that of a timelike boundary with a $\TTbar+\Lambda_2$ flow. 
A composite flow moves the screen across the cosmological horizon toward the static observer. 
These constructions suggest that different dS holographic descriptions correspond to different choices of boundary phase-space data.

\subsection{Flow equations in eigenvalue variables}\label{sec:solutions and flow eq}

For the Kerr-dS$_3$ application, we use an eigenvalue formulation of the flow equations ~\eqref{eq:flow eq for gamma} and \eqref{eq:flow eq for stress tensor}. 

Let $\gamma^{ab}_{[0]}$ denote the inverse of the undeformed metric $\gamma^{[0]}_{ab}$, and define
\begin{equation}
    M^a{}_b\equiv\gamma^{ac}_{[0]}\gamma_{cb}~.
\end{equation}
At $\lambda=0$,
\begin{equation}\label{eq:initialMT}
    M^a{}_b=\delta^a{}_b~,
    \qquad
    T^a{}_b=T^a_{[0]b}~.
\end{equation}

For the derivative-free class of deformations in \eqref{eq:class-function-O}, the flow is local in the boundary coordinates $x$: for position-dependent data, $x$ enters the characteristic equations only as a parameter. 
Since $\mathcal O$ is similarity invariant, its matrix derivative commutes with $T^a{}_b$. 
Together with \eqref{eq:initialMT}, this implies that $M(x,\lambda)$ and $T(x,\lambda)$ remain in the commutative algebra generated by $T_{[0]}(x)$. 
On a locally analytic branch, the Cayley--Hamilton theorem therefore gives
\begin{equation}\label{eq:pointwise-polynomial-flow}
    M(x,\lambda) = \sum_{k=0}^{d-1} m_k(x,\lambda)~T_{[0]}(x)^k~, \quad T(x,\lambda) = \sum_{k=0}^{d-1} \tau_k(x,\lambda)~T_{[0]}(x)^k~,
\end{equation}
and hence
\begin{equation}\label{eq:pointwise-commutators}
    [M,T_{[0]}] = [T,T_{[0]}] = [M,T] =0
\end{equation}
pointwise on the boundary.

For real symmetric initial stress-tensor data and a positive-definite Euclidean source metric, $T^a_{[0]b}(x)$ is self-adjoint and diagonalizable pointwise. 
On patches where its eigenvalue multiplicities are constant, a smooth local frame $S(x)$, independent of $\lambda$, simultaneously diagonalizes $T_{[0]}$, $M$, and $T$:
\begin{align}\label{eq:two matrix}
    S^{-1}T_{[0]}S = \mathrm{diag}\big(t_{\alpha,0}\big)~, \quad S^{-1}MS = \mathrm{diag}\big(\omega_\alpha\big)~, \quad S^{-1}TS = \mathrm{diag}\big(t_\alpha\big)~.
\end{align}
Thus the eigenspaces are preserved along the flow, although the diagonalizing frame may depend on $x$.

In eigenvalue variables, a deformation generated by $\mathcal O( T^a{}_b;\lambda)$ becomes a function of the eigenvalues,
\begin{equation}
	\mathcal O=\mathcal O(t_\alpha;\lambda)~.
\end{equation}
The metric-flow equations reduce to
\begin{align}
	\frac{\dd \omega_\alpha}{\dd\lambda} & =2\frac{\partial\mathcal O}{\partial t_\alpha}\omega_\alpha~,\label{eq:eigenvalue flow omega} \\
	\frac{\dd t_\alpha}{\dd\lambda} &= -\mathcal O + \sum^d_{\beta=1} t_\beta \frac{\partial\mathcal O}{\partial t_\beta} - t_\alpha \sum^d_{\beta=1} \frac{\partial\mathcal O}{\partial t_\beta}. \label{eq:eigenvalue flow t}
\end{align}
For position-dependent data, Eqs.~\eqref{eq:eigenvalue flow omega} and \eqref{eq:eigenvalue flow t} define ordinary differential equations in $\lambda$ at each fixed boundary point $x$. 
The diagonalizing matrix $S(x)$ should be understood as a local frame transformation rather than a coordinate transformation, and a smooth global eigenframe need not exist at degeneracy locations or on topologically nontrivial backgrounds. 
The invariant relations \eqref{eq:pointwise-polynomial-flow}--\eqref{eq:pointwise-commutators} are independent of this choice. 
For non-analytic deformations, such as root-$T\bar T$, one must in addition specify an analytic branch away from the discriminant locus. 
	
In the following, repeated eigenvalue indices are not summed over unless explicitly stated. 
As shown in App.~\ref{app:character II}, a useful corollary of the flow equations \eqref{eq:eigenvalue flow omega} and \eqref{eq:eigenvalue flow t} is 
\begin{equation}\label{eq:t t0 omega}
	\sqrt{\gamma}\frac{\pd t_\alpha}{\pd t_{\beta,0}}= \sqrt{\gamma_0}~\delta^\alpha_\beta+\frac{1}{2} \sqrt{\gamma_0} \sum^{d}_{\sigma=1} (t_{\sigma,0}-t_{\alpha,0}) \frac{\pd}{\pd t_{\beta,0}} \log \omega_\sigma~,
\end{equation}
where $\gamma_0=\gamma|_{\lambda=0}, t_{\alpha,0}\equiv t_\alpha|_{\lambda=0}$ denote their initial values at $\lambda=0$. 

We next restrict to stationary homogeneous operators, namely operators with no explicit $\lambda$ dependence and homogeneous of degree $m$ in the stress-tensor eigenvalues,
\begin{equation}\label{eq:def of order}
    \sum_{\alpha=1}^{d} t_\alpha\frac{\partial\mathcal O}{\partial t_\alpha} = m\mathcal O~.
\end{equation}
These assumptions are local in $x$ and do not require translationally invariant boundary data, so the solutions below apply pointwise. 
Eqs.~\eqref{eq:eigenvalue flow omega} and \eqref{eq:eigenvalue flow t} then imply
\begin{equation}\label{eq:sqrt gamma t-t const}
	\frac{\dd}{\dd\lambda}\left[\sqrt{\gamma}\mathcal{O}\right]=0~,
\end{equation}
and hence
\begin{equation}
	\sqrt{\gamma}\mathcal{O}(t_\alpha)=\sqrt{\gamma_0}\mathcal{O}(t_{\alpha,0})~.
\end{equation}
For $m\neq 1$, combining these relations gives
\begin{equation}\label{eq:solution t alpha}
	t_\alpha=\frac{t_{\alpha,0}+s}{\sqrt{\gamma/\gamma_0}}~,\quad s=\lambda(m-1)\mathcal O(t_{\alpha,0})~,
\end{equation}
with
\begin{equation}\label{eq:sqrt gamma solution}
	\sqrt{\frac{\gamma}{\gamma_0}}=\left[\frac{\mathcal O(t_{\alpha,0}+s)}{\mathcal{O}(t_{\alpha,0})}\right]^{\frac{1}{m-1}}~.
\end{equation}
Here $\mathcal O(t_{\alpha,0}+s)$ denotes $\mathcal O(t_{1,0}+s,\ldots,t_{d,0}+s)$.
For $m=1$, the degree-one eigenvalue flow is integrated directly.
An example is provided by the two-dimensional root-$\TTbar$ deformation \cite{Rodriguez:2021tcz,Conti:2022egv,Babaei-Aghbolagh:2022leo,Ferko:2022cix}, for which the present solutions reproduce the deformed boundary conditions discussed in \cite{Ebert:2023tih}.

Once $t_\alpha$ is known, the metric eigenvalues follow from
\begin{equation}\label{eq:solution omega alpha}
	\omega_\alpha(\lambda)=\exp\left[2\int_0^\lambda\dd\lambda'\frac{\partial\mathcal O}{\partial t_\alpha}\bigg|_{\lambda\to\lambda'}\right]~.
\end{equation}
This is the form of the mixed boundary condition used below.

The eigenvalue reduction above is local and remains valid pointwise for position-dependent data in any boundary dimension. 
This does not, however, guarantee a holographic completion of the flowed source--response data. 
For $d>2$, the state-dependent Fefferman--Graham response is encoded in $g^{(d)}_{ab}$ rather than generically in $g^{(2)}_{ab}$, and the expansion no longer truncates as in three bulk dimensions \cite{deHaro:2000vlm}. 
A higher-dimensional realization must therefore satisfy the bulk Einstein equations and constraints and admit an appropriate regular saddle; in even boundary dimensions, anomaly contributions must also be included. 
Higher-dimensional $T^2$ deformations provide examples of such large-$N$ constructions \cite{Hartman:2018tkw}. 
We therefore restrict the explicit holographic analysis below to $d=2$.

	
\section{Holography for dS$_3$/CFT$_2$ under deformations}\label{sec:deformed dS3}
	
Having derived the flow equations and solved the stationary homogeneous class, we now apply the construction to dS$_3$/CFT$_2$.
We compare the deformed boundary spectrum with that extracted from bulk data, thereby testing the proposed stress-tensor deformation dictionary.
	
	
\subsection{Deformed bulk Kerr-dS$_3$}
	
We begin with the Kerr-dS$_3$ geometry in the form
\begin{align}\label{eq:static patch in dS3 origin}
	\dd s^2 &= f(r)^{-1}\dd r^2-f(r)\dd t^2 + r^2\left(\dd \varphi-\frac{8G\mathcal{J}}{2r^2}\dd t \right)^2~,\\
	f(r) &= 8G\mathcal{M}-\frac{r^2}{\ell^2} +\frac{(8G\mathcal{J})^2}{4r^2}~. \notag
\end{align}
This parametrization makes stationarity and rotation manifest. 
Here $\mathcal M$ and $\mathcal J$ are the mass and angular-momentum parameters.
The locus $r=0$ encodes the nontrivial global identification and represents a spinning conical defect carrying these charges.

For later holographic use, it is convenient to rewrite \eqref{eq:static patch in dS3 origin} in terms of the cosmological-horizon parameter $r_+>0$ and rotation parameter $r_-$:
\begin{equation}\label{eq:intermediate dS}
    \dd s^2 = -\frac{\ell^2\tilde r^2}{(\tilde r^2+r_-^2)(\tilde r^2-r_+^2)}\dd\tilde r^2 +\frac{(\tilde r^2+r_-^2)(\tilde r^2-r_+^2)}{\tilde r^2}\dd x_1^2 +\tilde r^2\left(\dd x_2+\frac{r_+r_-}{\tilde r^2}\dd x_1\right)^2~,
\end{equation}
where
\begin{equation}
    \begin{split}
        \varphi &= \frac{2\pi}{R}x_2~,
        \qquad
        r = \frac{R}{2\pi}\tilde r~,
        \qquad
        t = \frac{2\pi\ell}{R}x_1~,
        \\
        \mathcal M &= \frac{R^2(r_+^2-r_-^2)}{32\pi^2G\ell^2}~,
        \qquad
        \mathcal J = -\frac{R^2r_+r_-}{16\pi^2G\ell}~.
    \end{split}
\end{equation}
The Kerr-dS identification $\varphi\sim\varphi+2\pi$ implies $x_2\sim x_2+R$. 
At $\mathcal I^+$ the leading boundary metric is Euclidean, and we identify $x_2$ as the compact spatial circle and $x_1$ as the noncompact Euclidean-time direction of the boundary cylinder. 
The latter is proportional to the stationary coordinate $t$, although $\partial_t$ is not globally timelike. 
The asymptotic region is at $\tilde r\to\infty$, while the cosmological horizon is located at $\tilde r=r_+$.\footnote{For $r_-=0$ and $r_+=2\pi\ell/R$, the conical defect disappears and the metric reduces to the pure-dS$_3$ static patch.}

Using the coordinate transformations
\begin{equation}\label{eq:r rho z x}
	\tilde r^2 = \frac{1}{\rho }+\frac{1}{16} \rho  \left(r_-^2+r_+^2\right){}^2+\frac{1}{2} \left(r_+^2-r_-^2\right), \quad z =  x_2 + i x_1~, \quad \bar z =  x_2 - i x_1~,
\end{equation}
the rescaled metric \eqref{eq:intermediate dS} takes Fefferman--Graham form \cite{fefferman1985conformal}:
\begin{equation}\label{eq:FG for dS3}
	\dd s^2 = -\frac{\ell^2}{4\rho^2} \dd\rho^2 + \frac{1}{\rho} \dd z \dd \bar z + \mathcal{L} \dd z^2 + \bar{\mathcal L} \dd \bar z^2 + \rho \mathcal{L} \bar{\mathcal L} \dd z \dd \bar z~,
\end{equation}    
where
\begin{equation}\label{eq:L Lb r+-}
	\mathcal{L} = \frac{1}{4} \left(r_+-i r_-\right){}^2,~~~ \mathcal{\bar L}= \frac{1}{4} \left(r_++i r_-\right){}^2~.
\end{equation}
This geometry is the dS$_3$ analogue of the Ba\~nados family in AdS$_3$ \cite{Banados:1992wn,Banados:1992gq,Banados:1998gg}. The conformal boundary $\mathcal I^+$ is at $\rho\to0^+$, and the cosmological horizon is located at
\begin{equation}
	\rho_h = (\mathcal{L}\bar{\mathcal L})^{-1/2}.
\end{equation}

For the Fefferman--Graham expansion
\begin{equation}\label{eq:FG gauge}
	\dd s^2 = -\frac{\ell^2}{4\rho^2}~\dd\rho^2 + \frac{1}{\rho} \left( g^{(0)}_{ab} + \rho g^{(2)}_{ab} + \rho^2 g^{(4)}_{ab} \right)\dd x^a\dd x^b~,
\end{equation}
holographic renormalization gives the boundary source and response as
\begin{equation}\label{eq:holographic-renorm}
	\gamma_{ab}=g^{(0)}_{ab}~,
	\qquad
	T_{ab}=T^{\rm BY}_{ab}~,
\end{equation}
with the Brown-York tensor \cite{Brown:1992br,Henningson:1998gx,Skenderis:1999nb,deHaro:2000vlm}
\begin{equation}\label{eq:Brown-York-in-dS3}
	T^{\rm BY}_{ab} = \frac{1}{8\pi G\ell} \left( g^{(2)}_{ab} - g^{(2)c}{}_{c}~g^{(0)}_{ab} \right)~.
\end{equation}
For the Kerr-dS$_3$ geometry, Eqs.~\eqref{eq:r rho z x}, \eqref{eq:FG for dS3}, and \eqref{eq:L Lb r+-} therefore give the undeformed data
\begin{equation}
	\gamma^{[0]}_{ab} =
	\begin{pmatrix}
		1 & 0\\
		0 & 1
	\end{pmatrix}~,
	\qquad
	T^{[0]}_{ab}
	=
	\frac{1}{4\pi G\ell}
	\begin{pmatrix}
		\dfrac{r_-^2-r_+^2}{4} & \dfrac{r_-r_+}{2}\\
		\dfrac{r_-r_+}{2} & \dfrac{r_+^2-r_-^2}{4}
	\end{pmatrix}~.
\end{equation}

The boundary Killing vectors $\partial_{x_1}$ and $\partial_{x_2}$ define the corresponding Brown--York charges
\begin{align}
	E^{[0]} = & -\int_0^R\dd x_2~T^{1}{}_{1,0} = \frac{R}{16\pi G\ell} \left(r_+^2-r_-^2\right)~, \label{eq:undeformed energy} \\
	J^{[0]} = & -\int_0^R\dd x_2~T^{1}{}_{2,0} = -\frac{R}{8\pi G\ell}~r_+r_-~. \label{eq:undeformed angular}
\end{align}
Their conservation follows from the boundary Ward identity $\nabla^aT_{ab}=0$. 
Since $\mathcal I^+$ is spacelike, $E^{[0]}$ should not be interpreted as a Lorentzian bulk energy generated by a globally timelike Killing vector. 
Rather, in the putative Euclidean dual, $E^{[0]}$ and $J^{[0]}$ generate translations along the noncompact $x_1$ direction and the compact $x_2$ circle, respectively. 
Their normalization relative to the bulk parameters $\mathcal M$ and $\mathcal J$ follows from the coordinate transformation above.

For comparison, the Kerr-dS$_3$ parametrization is related to the Ba\~nados--Teitelboim--Zanelli (BTZ) black hole by the continuation \eqref{eq:AdS to dS continuation}, together with $\mathcal M\to-\mathcal M$ \cite{Banados:1992wn,Banados:1992gq,Park:1998qk,Hikida:2022ltr,Balasubramanian:2001nb}. 
This accounts for the relative sign between the $r_+^2$ and $r_-^2$ terms in \eqref{eq:undeformed energy}.

To apply the eigenvalue flow of Sec.~\ref{sec:solutions and flow eq}, we work in a frame that diagonalizes the undeformed Kerr-dS$_3$ boundary data. 
Since $\mathcal L$ and $\bar{\mathcal L}$ are constant, the local eigenframe can be realized globally by the linear transformation
\begin{equation}
	z = \left( \frac{\bar{\mathcal L}}{\mathcal L} \right)^{1/4} (x+i\tau)~, \qquad \bar z = \left( \frac{\mathcal L}{\bar{\mathcal L}} \right)^{1/4} (x-i\tau)~.
\end{equation}
In these coordinates, the Fefferman--Graham metric \eqref{eq:FG for dS3} becomes
\begin{equation}\label{eq:diagonal FG}
	\dd s^2 = -\frac{\ell^2}{4\rho^2}~\dd\rho^2 + \frac{\dd\tau^2+\dd x^2}{\rho} + 2\sqrt{\mathcal L\bar{\mathcal L}} \left( -\dd\tau^2+\dd x^2 \right) + \rho\mathcal L\bar{\mathcal L} \left( \dd\tau^2+\dd x^2 \right)~.
\end{equation}
For generic position-dependent data, such a global coordinate realization need not exist; the pointwise eigenframe construction of Sec.~\ref{sec:solutions and flow eq} remains applicable. 
The corresponding undeformed stress-tensor eigenvalues are
\begin{equation}\label{eq:initial values of ti0}
	t_{1,0} = -t_{2,0}  = -h~, \qquad h \equiv \frac{\sqrt{\mathcal L\bar{\mathcal L}}}{4\pi G\ell}~.
\end{equation}

The bulk metric \eqref{eq:diagonal FG} solves the undeformed de Sitter Einstein equations. 
The deformation changes only the boundary condition, which selects the metric and response tensor of the deformed boundary theory. 
Applying the general flow solution of Sec.~\ref{sec:solutions and flow eq} gives the diagonal-frame boundary metric
\begin{equation}\label{eq:deformed bdry metric}
	\gamma_{ab}\dd x^a \dd x^b = \omega_1 \dd \tau^2 + \omega_2 \dd x^2~.
\end{equation}
To place the deformed theory on a standard Euclidean cylinder of circumference $R$, we introduce deformation- and state-dependent canonical coordinates,
\begin{equation}\label{eq:tau x to T phi}
	\tau=\frac{1}{\sqrt{\omega_1}}(T \cos\theta + \phi \sin\theta)~,\qquad x = \frac{1}{\sqrt{\omega_2}} (-T \sin\theta + \phi \cos\theta)~,
\end{equation}
and impose
\begin{equation}
	\gamma_{ab} \dd x^a \dd x^b=\dd T^2 + \dd \phi^2~,\qquad\phi\sim\phi+R~.
\end{equation}
Here $(T,\phi)$ are the canonical field-theory coordinates. 
At finite deformation, this is not a coordinate rewriting of a fixed global Kerr-dS$_3$ quotient.
It defines a locally de Sitter saddle whose boundary conditions and global identifications vary so that the boundary theory remains on the prescribed cylinder. 
The undeformed Kerr-dS$_3$ quotient is recovered as the initial condition at $\lambda=0$.
	
Expressed in the canonical coordinates, the corresponding bulk saddle is\footnote{For $\lambda_1=0$, \eqref{eq:deformed dS3 bulk geometry} is obtained from the Guica--Monten metric \cite{Guica:2019nzm} (their Eqs.~(2.28) and (3.6)--(3.7)) by $\ell_{\rm AdS}\to-i\ell$ together with the Wick rotation of the AdS boundary time, on the branch fixed in \eqref{eq:AdS to dS continuation}.}
\begin{align}\label{eq:deformed dS3 bulk geometry}
	\dd s^2 = & -\frac{\ell^2}{4\rho^2} \dd\rho^2 + \frac{\omega_2 \cos^2\theta (1 - \rho \sqrt{\mathcal{L} \bar{\mathcal L}})^2 + \omega_1 \sin^2\theta (1 + \rho \sqrt{\mathcal{L} \bar{\mathcal L}})^2}{\rho \omega_1 \omega_2} \dd T^2 \notag \\
	& + \frac{\omega_2 \sin^2\theta (1 - \rho \sqrt{\mathcal{L} \bar{\mathcal L}})^2 + \omega_1 \cos^2\theta (1 + \rho \sqrt{\mathcal{L} \bar{\mathcal L}})^2}{\rho \omega_1 \omega_2} \dd \phi^2 \notag \\
	&+ \frac{\sin 2\theta \Big[ \omega_2 (1 - \rho \sqrt{\mathcal{L} \bar{\mathcal L}})^2 - \omega_1 (1 + \rho \sqrt{\mathcal{L} \bar{\mathcal L}})^2 \Big]}{\rho \omega_1 \omega_2} \dd T \dd \phi~,
\end{align}
while the deformed boundary stress tensor becomes
\begin{align}\label{eq:deformed dS3 stress tensor general}
	T_{ab} \dd x^a \dd x^b =& (t_1 \cos^2\theta + t_2 \sin^2\theta) \dd T^2 + (t_1 \sin^2\theta + t_2 \cos^2\theta) \dd \phi^2 \notag \\
	& + \sin 2\theta (t_1 - t_2) \dd T \dd \phi~.
\end{align}
Equations~\eqref{eq:deformed dS3 bulk geometry} and \eqref{eq:deformed dS3 stress tensor general} thus realize the deformed boundary theory in the bulk. 
We next compare the spectrum extracted from this geometry with the field-theory flow.
	
	
\subsection{Matching of the spectrum}\label{sec:match spectrum}

The deformed boundary stress tensor \eqref{eq:deformed dS3 stress tensor general} defines the charges
\begin{align}
	E^{[\lambda]} &= -\int_0^R \dd\phi~ T^{T}{}_{T}
	= -R\left(t_1\cos^2\theta + t_2\sin^2\theta\right)~, \label{eq:deformed energy}\\
	J^{[\lambda]} &= -\int_0^R \dd\phi~ T^{T}{}_{\phi}
	= -\frac{R}{2}\left(t_1 - t_2\right)\sin 2\theta~. \label{eq:deformed angular}
\end{align}
To clarify the status of the spectrum comparison, the mixed boundary condition defined by Eqs.~\eqref{eq:flow eq for gamma}--\eqref{eq:flow eq for stress tensor} is taken as the holographic prescription rather than derived from the Fefferman--Graham expansion or the Brown--York relation. 
The undeformed Kerr-dS$_3$ source and Brown--York response provide the initial data, while the fixed-state conditions \eqref{eq:J constraint}--\eqref{eq:A constraint}, introduced below, supply independent global information. 
The deformed Brown--York charges \eqref{eq:deformed energy} and \eqref{eq:deformed angular} are then determined outputs. 

In pure three-dimensional Einstein gravity, the Fefferman--Graham expansion truncates as
$g^{(4)}=\frac{1}{4}~g^{(2)}(g^{(0)})^{-1}g^{(2)}$, so imposing the asymptotic data and the mixed source--response relation leaves no additional independent local bulk equation. 
The spectrum comparison should therefore be regarded as an internal consistency check rather than an independent derivation from bulk dynamics. 
Its nontrivial content lies in the compatibility of the local flow with the global quotient and state identification: the momentum and horizon-area constraints, together with the standard-cylinder condition, determine $h(\lambda,R)$ and $\theta(\lambda,R)$. 
The agreement is consequently sensitive to the Brown--York normalization and to the consistency of these global conditions with the local source--response flow.

Using $\rho_h = (\mathcal{L}\bar{\mathcal L})^{-1/2}$, the induced metric on the cosmological horizon takes the form
\begin{equation}
	\dd s^2|_{\rho=\rho_h}=\frac{4\sqrt{\mathcal{L} \bar{\mathcal L}}\sin^2\theta}{\omega_2}\dd T^2+\frac{4\sqrt{\mathcal{L} \bar{\mathcal L}}\cos^2\theta}{\omega_2}\dd\phi^2 -\frac{4\sqrt{\mathcal{L} \bar{\mathcal L}}\sin2\theta}{\omega_2}\dd T\dd\phi~,
	\end{equation}
from which the horizon area is
\begin{equation}\label{eq:deformed area}
	A^{[\lambda]} \equiv \int_0^R \dd\phi \sqrt{g_{\phi\phi}} = \frac{2R\sqrt{4\pi G \ell}\sqrt{h}\cos\theta}{\sqrt{\omega_2}}~.
\end{equation}

Having determined the deformed charges and horizon area, we now specify how a state is followed as $\lambda$ and the circumference $R$ vary. 
Following the fixed-state prescription of \cite{Guica:2019nzm,Ebert:2023tih}, we keep fixed the dimensionless momentum combination $JR$ and the cosmological-horizon area. 
The former follows from momentum quantization on the compact circle, while the latter is motivated by the identification of the horizon area with the state degeneracy. 
These conditions are independent holographic inputs and are not fixed by the local metric flow or by the AdS-to-dS continuation of Sec.~\ref{subsec:ads-to-ds}. 
We therefore impose
\begin{align}
	C_{J}
	&\equiv
	\frac{R^2 h\sin 2\theta}{\sqrt{\gamma/\gamma_0}}
	=
	\text{constant}~,
	\label{eq:J constraint}\\
	C_{A}
	&\equiv
	\frac{R\sqrt{h}\cos\theta}{\sqrt{\omega_2}}
	=
	\text{constant}~.
	\label{eq:A constraint}
\end{align}
In \eqref{eq:J constraint} we used \eqref{eq:initial values of ti0} and \eqref{eq:sqrt gamma t-t constant}. 
Both constants are fixed by the undeformed data. 
Let $h_0$ and $\theta_0$ characterize the undeformed geometry. 
At $\lambda=0$, one has $\omega_1=\omega_2=1$ and $\gamma/\gamma_0=1$, and the undeformed energy and angular momentum become 
\begin{equation}
	E^{[0]} = R h_0 \cos 2\theta_0~, \quad J^{[0]} = R h_0 \sin 2\theta_0~,
\end{equation}
where we used \eqref{eq:initial values of ti0}. 
Thus
\begin{equation}
	\sqrt{(E^{[0]})^2+(J^{[0]})^2}=R h_0~,
\end{equation}
so the constants can be written as
\begin{align}
	C_J &= R J^{[0]}~,\\
	C_A &= R\sqrt{h_0}\cos\theta_0 = \sqrt{\frac{R}{2}\left(\sqrt{(E^{[0]})^2+(J^{[0]})^2}+E^{[0]}\right)}~.
\end{align}
The constraints \eqref{eq:J constraint} and \eqref{eq:A constraint} can then be written directly in terms of the undeformed spectrum:
\begin{align}
	\sin 2\theta &= \frac{J^{[0]}\sqrt{\omega_1\omega_2}}{hR}~,\label{eq:newCJ} \\
	2hR &= \sqrt{(E^{[0]})^2+(J^{[0]})^2}\,(\omega_1+\omega_2) +E^{[0]}(\omega_2-\omega_1)~. \label{eq:newCA}
\end{align}
After solving for $\omega_1(\lambda,h)$ and $\omega_2(\lambda,h)$, Eqs.~\eqref{eq:newCJ} and \eqref{eq:newCA} determine $h$ and $\theta$ from $(\lambda,R)$ and the undeformed spectrum. 
The derivation below requires only that $C_J$ and $C_A$ be constant, while \eqref{eq:newCJ}--\eqref{eq:newCA} show explicitly how the undeformed boundary data fix the bulk integration constants. 

Treating $\omega_\alpha$ and $t_\alpha$ as functions of $(\lambda,h)$, we write
\begin{equation}\label{eq:domega and dt}
	\dd\omega_\alpha=\pd_\lambda \omega_\alpha \dd\lambda+\pd_h \omega_\alpha\dd h~,\quad \dd t_\alpha=\pd_\lambda t_\alpha\dd\lambda+\pd_h t_\alpha\dd h~.
\end{equation}
Equations~\eqref{eq:eigenvalue flow omega} and \eqref{eq:eigenvalue flow t} determine their $\lambda$ dependence:
\begin{align}
	\pd_\lambda \omega_\alpha=& 2\frac{\pd\mathcal{O}}{\pd t_\alpha}\omega_\alpha~,\\
	\pd_\lambda t_1=& \frac{2h}{\sqrt{\gamma/\gamma_0}} \frac{\pd\mathcal{O}}{\pd t_2} -\mathcal{O}~,\\
	\pd_\lambda t_2=& \frac{-2h}{\sqrt{\gamma/\gamma_0}}\frac{\pd\mathcal{O}}{\pd t_1}-\mathcal{O}~,
\end{align}
where Eqs.~\eqref{eq:initial values of ti0} and \eqref{eq:sqrt gamma t-t constant} have been used. 
In terms of the initial values, 
\begin{equation}
	\frac{\pd}{\pd h}=-\frac{\pd}{\pd t_{1,0}}+\frac{\pd}{\pd t_{2,0}}~,
\end{equation}
which, together with Eq.~\eqref{eq:t t0 omega}, implies the $h$-dependence
\begin{equation}
	\sqrt{\gamma}\pd_h t_1=-\sqrt{\gamma_0}+\sqrt{\gamma_0} h\pd_h\log\omega_2~,\quad \sqrt{\gamma}\pd_h t_2=\sqrt{\gamma_0}-\sqrt{\gamma_0}h\pd_h\log\omega_1~.
\end{equation}
The constraint
\begin{equation}
	\dd\left(\frac{C_J}{C_A}\right)\equiv 0 
\end{equation}
then implies that
\begin{equation}\label{eq:expression for dtheta}
	\dd\theta=\tan\theta\frac{\pd\mathcal{O}}{\pd t_1} \dd\lambda+\frac{\tan\theta}{2h}(h\pd_h\log\omega_1-1)\dd h-\frac{\tan\theta}{R}\dd R~.
\end{equation} 
Substituting this result into the constraint
\begin{equation}
	\dd C_A\equiv0~,
\end{equation}
we obtain
\begin{equation}\label{eq:solution for dh}
	\dd h = \frac{4 h \left(R \left(\frac{\pd\mathcal{O}}{\pd t_1} \sin ^2\theta +\frac{\pd\mathcal{O}}{\pd t_2} \cos ^2 \theta \right)\dd\lambda - \dd R\right)}{R [ h \cos 2 \theta (\pd_h\log\omega_1-\pd_h\log\omega_2)-h (\pd_h\log\omega_1+\pd_h\log\omega_2)+2]}~.
\end{equation}
Equations~\eqref{eq:domega and dt}, \eqref{eq:expression for dtheta}, and \eqref{eq:solution for dh} then give
\begin{align}
	\dd E^{[\lambda]}=& \pd_R E^{[\lambda]}\dd R+\pd_{\theta}E^{[\lambda]}\dd\theta+\pd_{t_1}E^{[\lambda]} \dd t_1+\pd_{t_2}E^{[\lambda]}\dd t_2 \notag\\
	=& R \mathcal{O}\dd\lambda- (t_1\sin^2\theta+t_2\cos^2\theta) \dd R~,
\end{align}
which implies that the deformed spectrum satisfies
\begin{equation}\label{eq:flow on spectrum}
	\frac{\pd E^{[\lambda]}}{\pd\lambda}=\int^R_0 \mathcal{O} \dd\phi ~,\quad \frac{\pd E^{[\lambda]}}{\pd R}=- T_{\phi\phi}~.
\end{equation}
Thus the deformed geometry \eqref{eq:deformed dS3 bulk geometry} and stress tensor \eqref{eq:deformed dS3 stress tensor general} reproduce the field-theory spectral flow.
The comparison is nontrivial because the bulk construction fixes $h$ and $\theta$ through horizon-area and momentum constraints, whereas the field-theory flow uses only the deformed stress tensor on the standard cylinder.


\section{Pseudo entropy}\label{sec:EE}

We next study a nonlocal probe of the deformed saddle. 
Boundary-anchored geodesics in dS$_3$ are generically complex. 
Following the proposal that complex de Sitter extremal surfaces encode pseudo entropy \cite{Nakata:2020luh,Mollabashi:2020yie,Mollabashi:2021xsd,Doi:2022iyj,Narayan:2015vda,Narayan:2015oka}, we refer to the corresponding complex-geodesic observable as the holographic pseudo entropy of a boundary interval. 
For generic stress-tensor-deformed Kerr-dS$_3$ intervals, this identification is a semiclassical prescription whose finite-$n$ replica derivation remains open.
For the stress-tensor-deformed theory, we adopt the same prescription and discuss its wavefunction-replica interpretation and a perturbative boundary check below and in App.~\ref{app:replica pseudo entropy}. 
	
dS$_3$ is the hyperboloid
\begin{equation}\label{eq:embedding dS3}
	-(X^0)^2+(X^1)^2+(X^2)^2+(X^3)^2=\ell^2~,
\end{equation}
embedded in four-dimensional Minkowski space $\mathbb{R}^{1,3}$ with metric
\begin{equation}
	\dd s^2=-(\dd X^0)^2+(\dd X^1)^2+(\dd X^2)^2+(\dd X^3)^2~.
\end{equation}
For the deformed bulk geometry \eqref{eq:deformed dS3 bulk geometry}, a convenient parametrization is
\begin{align}
	X^0=& \frac{\ell}{2}\left(\frac{1}{\mathcal{N}\sqrt{\rho}}-\mathcal{N}\sqrt{\rho}\right)\cosh\left(\frac{2\mathcal{N}}{\ell}\frac{1}{\sqrt{\omega_1}} (T\cos\theta+\phi\sin\theta)\right)~,\label{eq:embedding X3}\\
	X^3=& \frac{\ell}{2}\left(\frac{1}{\mathcal{N}\sqrt{\rho}}-\mathcal{N}\sqrt{\rho}\right)\sinh\left(\frac{2\mathcal{N}}{\ell}\frac{1}{\sqrt{\omega_1}} (T\cos\theta+\phi\sin\theta)\right)~,\\
	X^1=& \frac{\ell}{2}\left(\frac{1}{\mathcal{N}\sqrt{\rho}}+\mathcal{N}\sqrt{\rho}\right)\cos\left(\frac{2\mathcal{N}}{\ell}\frac{1}{\sqrt{\omega_2}}(-T\sin\theta+\phi\cos\theta)\right)~,\\
	X^2=& \frac{\ell}{2}\left(\frac{1}{\mathcal{N}\sqrt{\rho}}+\mathcal{N}\sqrt{\rho}\right)\sin\left(\frac{2\mathcal{N}}{\ell}\frac{1}{\sqrt{\omega_2}}(-T\sin\theta+\phi\cos\theta)\right)~,\label{eq:embedding X2}
\end{align}
where
\begin{equation}\label{eq:def for cN}
	\mathcal{N}^2\equiv \sqrt{\mathcal{L}\bar{\mathcal L}}=4\pi G \ell h~,
\end{equation}
and we take $\mathcal N>0$. The geodesic distance $d$ between two points $X_\alpha$ and $X_\beta$ obeys
\begin{equation}
	\cos\left(\frac{d}{\ell}\right)
		=\frac{-X^0_\alpha X^0_\beta+X^1_\alpha X^1_\beta+X^2_\alpha X^2_\beta+X^3_\alpha X^3_\beta}{\ell^2}~.
\end{equation}
Consider an interval on the standard boundary cylinder at fixed $T=T_0$, with endpoints
\begin{equation*}
	(\rho,T,\phi)=(\rho_0,T_0,0)~,\quad \text{and} \quad  (\rho_0,T_0,\phi_0)~.
\end{equation*}
We take $0<\phi_0<R$.
Because $\phi\sim\phi+R$, the second endpoint has lifts
\begin{equation}\label{eq:interval-image-lifts-phi}
	\phi_n\equiv\phi_0+nR~,\qquad n\in\mathbb Z~.
\end{equation}
Let $d_n$ denote the geodesic distance to the $n$th lift. 
Translation invariance lets us set $T_0=0$. Substituting the endpoints into the embedding invariant gives
\begin{align}
	\cos\left(\frac{d_n}{\ell}\right)=&\frac{1}{4}\left(\frac{1}{\mathcal{N}\sqrt{\rho_0}}+\mathcal{N}\sqrt{\rho_0}\right)^2 \cos\left(\frac{2\mathcal{N}\phi_n\cos\theta}{\ell\sqrt{\omega_2}}\right)\notag\\
    &-\frac{1}{4}\left(\frac{1}{\mathcal{N}\sqrt{\rho_0}}-\mathcal{N}\sqrt{\rho_0}\right)^2\cosh\left(\frac{2\mathcal{N}\phi_n\sin\theta}{\ell\sqrt{\omega_1}}\right)~.\label{eq:cos geodesic deformed}
\end{align}
Equation~\eqref{eq:cos geodesic deformed} determines only $\cos(d_n/\ell)$, leaving two independent choices: the image label $n$ and the inverse-cosine sheet for each $d_n$. 
For generic rotation, we restrict to the direct image $n=0$ and choose the principal inverse-cosine sheet. 
These choices specify the saddle used below and are not fixed by the local AdS-to-dS continuation of Sec.~\ref{subsec:ads-to-ds}. 
A complete comparison of rotating image saddles would require additional information about the complexified gravitational integration cycle.

For $\theta=0$, App.~\ref{app:complex-geodesic-nonrotating} compares smooth real geodesics in an auxiliary hyperbolic cone with opening angle
\begin{equation}
    U\equiv\frac{2\mathcal N R}{\ell\sqrt{\omega_2}}~.
    \label{eq:main-cone-angle}
\end{equation}
We select the complex geodesic by continuing the shortest smooth connected auxiliary geodesic and independently choosing the principal inverse-cosine sheet. 
In the no-conical-excess sector, $0<U\leq2\pi$, the shortest connected auxiliary image is 
\begin{equation}\label{eq:shortest-image-label}
	n_\star(\phi_0)
	=
	\begin{cases}
		0~,  & 0<\phi_0<R/2~,\\[2mm]
		-1~, & R/2<\phi_0<R~.
	\end{cases}
\end{equation}
Within this sector, the selected connected result for $0<\phi_0<R$ is obtained by replacing 
\begin{equation}\label{eq:shortest-interval-size}
	\phi_0
	\longrightarrow
	\widehat\phi_0
	\equiv
	\min\{\phi_0,R-\phi_0\}~.
\end{equation}
At $\phi_0=R/2$, the two auxiliary images have equal length.
Under this prescription, their exchange produces a cusp in the connected pseudo entropy for $0<U<2\pi$. 
For $U=2\pi$, they represent the same unoriented diameter, and the result is smooth.
This cusp characterizes the selected connected contribution; establishing a transition in the full replica functional requires the contributing saddles and their weights. 
Along the fixed-state flow of Sec.~\ref{sec:match spectrum}, the horizon-area constraint fixes $U$, preserving this sector and the exchange point $\phi_0/R=1/2$. 
This invariance does not apply when the state is varied at fixed deformation parameters.

Unless stated otherwise, the formulae below refer to the direct image $n=0$, and we write $d=d_0$. 
For non-rotating intervals with $\phi_0>R/2$, the connected result is understood through the replacement \eqref{eq:shortest-interval-size}.

In the boundary limit $\rho_0=(\epsilon/\ell)^2\to0$, the argument of \eqref{eq:cos geodesic deformed} is large and negative on the selected saddle. 
Using $\arccos(-|X|)=\pi-i\ln(2|X|)+\mathcal O(|X|^{-2})$ on the principal inverse-cosine sheet, we obtain
\begin{equation}\label{eq:general geodesic distance}
	d
	\approx
	\pi\ell
	-i\ell\ln\left[
	\frac{\ell^2}{\mathcal N^2\epsilon^2}
	\left(
	\sinh^2\frac{\mathcal N\phi_0\sin\theta}{\ell\sqrt{\omega_1}}
	+
	\sin^2\frac{\mathcal N\phi_0\cos\theta}{\ell\sqrt{\omega_2}}
	\right)
	\right]~.
\end{equation}
Following the complex-extremal-surface prescription
\cite{Mollabashi:2020yie,Mollabashi:2021xsd,Doi:2022iyj,Narayan:2015vda,Narayan:2015oka},
we associate to the selected geodesic the semiclassical pseudo entropy
\begin{equation}\label{eq:general pseudo entropy}
	S^{[\lambda]}(\phi_0)
	=
	\frac{d}{4G}
	=
	\frac{\pi c_{\rm dS}}{6}
	-\frac{i c_{\rm dS}}{6}
	\ln\left[
	\frac{\ell^2}{\mathcal N^2\epsilon^2}
	\left(
	\sinh^2\frac{\mathcal N\phi_0\sin\theta}{\ell\sqrt{\omega_1}}
	+
	\sin^2\frac{\mathcal N\phi_0\cos\theta}{\ell\sqrt{\omega_2}}
	\right)
	\right]~,
\end{equation}
where $c_{\rm dS}=3\ell/(2G)$ is the real dS central charge in the Brown--York convention introduced in Sec.~\ref{subsec:ads-to-ds}. 
On the principal inverse-cosine sheet, the additive real term $\pi c_{\rm dS}/6$ is numerically equal to one half of the Gibbons--Hawking entropy of pure dS$_3$ \cite{Gibbons:1977mu}. 
On the non-rotating principal branch, the leading large-$c$ replica calculation in App.~\ref{app:replica pseudo entropy} reproduces the interval-dependent $O(\lambda_1,\lambda_2)$ corrections. 
The remaining interval-independent difference is absorbed by the source-frame regulator matching specified there. 

Define
\begin{equation}
	\alpha_\pm
	\equiv
	\frac{\mathcal N}{\ell}
	\left(
	\frac{\cos\theta}{\sqrt{\omega_2}}
	\pm i\frac{\sin\theta}{\sqrt{\omega_1}}
	\right)~.
\end{equation}
Equation~\eqref{eq:general pseudo entropy} can then be written as
\begin{equation}\label{eq:general pseudo entropy chiral}
	S^{[\lambda]}(\phi_0)
	=
	\frac{\pi c_{\rm dS}}{6}
	-\frac{i c_{\rm dS}}{6}
	\log\left[
	\frac{\ell^2}{\mathcal N^2\epsilon^2}
	\sin(\alpha_+\phi_0)\sin(\alpha_-\phi_0)
	\right]~.
\end{equation}
For the real parameters considered here, $\alpha_-=\alpha_+^*$, so the logarithm has a positive argument away from its zeros and introduces no additional branch ambiguity.

For vanishing momentum, $J^{[0]}=0$, we choose the branch
\begin{equation}
	\theta=0~,
\end{equation}
of Eq.~\eqref{eq:newCJ}. 
The pseudo entropy \eqref{eq:general pseudo entropy} then reduces to
\begin{equation}\label{eq:nonrotating pseudo entropy}
	S^{[\lambda]}(\phi_0)
	=
	\frac{\pi c_{\rm dS}}{6}
	-\frac{i c_{\rm dS}}{3}
	\ln\left[
	\frac{\ell}{\mathcal N\epsilon}
	\sin\left(
	\frac{\mathcal N\phi_0}{\ell\sqrt{\omega_2}}
	\right)
	\right]~.
\end{equation}
We define the inverse horizon-regularity scale measured in the canonical Euclidean time $T$ by
\begin{equation}\label{eq:effective temperature general}
	\beta \equiv \frac{1}{\mathscr T_{\rm dS}} = \frac{\pi\ell\sqrt{\omega_1}}{\mathcal N}~.
\end{equation}
For the rotating saddle, the corresponding period is fixed by smoothness of the contractible Euclidean cycle generated by the co-rotating horizon Killing field; Eq.~\eqref{eq:effective temperature general} is its $\theta=0$ specialization, as derived in App.~\ref{app:Temperature}. 
Thus $\beta$ is a horizon-regularity scale in canonical boundary-cylinder units, rather than an assumed Gibbs inverse temperature of a unitary Lorentzian boundary theory. 
Equation~\eqref{eq:nonrotating pseudo entropy} becomes
\begin{equation}\label{eq:nonrotating pseudo entropy beta}
	S^{[\lambda]}(\phi_0)
	=
	\frac{\pi c_{\rm dS}}{6}
	-\frac{i c_{\rm dS}}{3}
	\ln\left[
	\frac{\beta}{\pi\sqrt{\omega_1}\epsilon}
	\sin\left(
	\frac{\pi\sqrt{\omega_1}\phi_0}
	{\beta\sqrt{\omega_2}}
	\right)
	\right]~.
\end{equation}
The factors of $\omega_1$ and $\omega_2$ encode the conversion between the diagonal flow frame and the canonical boundary cylinder.
	
In the undeformed limit, \eqref{eq:nonrotating pseudo entropy beta} gives
\begin{equation}\label{eq:undeformed HEE}
	S^{[0]}(\phi_0)
		=\frac{\pi c_{\rm dS}}{6}
		-\frac{i c_{\rm dS}}{3}
		\ln\left[
		\frac{\beta_0}{\pi\epsilon}
		\sin\left(\frac{\pi\phi_0}{\beta_0}\right)
		\right]~,
\end{equation}
where $\beta_0=\pi\ell/\mathcal{N}_0$. 
This agrees with the usual dS$_3$/CFT$_2$ result obtained by analytic continuation from the corresponding AdS$_3$ geodesic formula.


\subsection{$\TTbar$ and root-$\TTbar$ deformed pseudo entropy}\label{sec:TT entropy}

We now specialize to the commuting $\TTbar$ and root-$\TTbar$ flows in two dimensions. 
Throughout this subsection, $t_\alpha\equiv t_{\alpha,{\rm BY}}$ denotes an eigenvalue of the real Brown--York response tensor, while $\lambda_1,\lambda_2\in\mathbb R$ are the mixed-boundary-condition parameters. 
In the normalization of the real source--response flow, we take
\begin{equation}
	\mathcal O_{\rm BY}^{(2)}
	=
	t_1t_2~,
	\qquad
	\mathcal O_{\rm BY}^{(1)}
	=
	t_2-t_1~,
	\label{eq:TTroot operators BY}
\end{equation}
corresponding respectively to the $\TTbar$ deformation and to the chosen branch of the root-$\TTbar$ deformation
\cite{Zamolodchikov:2004ce,Smirnov:2016lqw,Cavaglia:2016oda,Rodriguez:2021tcz,Conti:2022egv,Babaei-Aghbolagh:2022leo,Ferko:2022cix}.

To translate these deformations to the wavefunction convention of Sec.~\ref{subsec:ads-to-ds}, let $t_{\alpha,\Psi}$ denote the eigenvalues of $T_{\Psi}=-iT_{\rm BY}$, so that
$t_\alpha=i t_{\alpha,\Psi}$. 
We choose the corresponding operator normalizations
\begin{equation}
	\widehat{\mathcal O}_{\Psi}^{(2)} \equiv -t_{1,\Psi}t_{2,\Psi}~, \qquad \widehat{\mathcal O}_{\Psi}^{(1)} \equiv t_{2,\Psi}-t_{1,\Psi}~. 
\end{equation}
On the selected root-$\TTbar$ branch,
\begin{equation}
	\mathcal O_{\rm BY}^{(2)} = \widehat{\mathcal O}_{\Psi}^{(2)}~, \qquad \mathcal O_{\rm BY}^{(1)} = i\widehat{\mathcal O}_{\Psi}^{(1)}~.
\end{equation}
Using $I_{\Psi}=-iS_{\rm ren}^{\rm dS}$, the real mixed-boundary-condition flow therefore becomes
\begin{equation}
	\pd_{\lambda_2}I_{\Psi} = -i\int\dd^2x~\sqrt\gamma~ \widehat{\mathcal O}_{\Psi}^{(2)}~, \qquad \pd_{\lambda_1}I_{\Psi} = \int\dd^2x~\sqrt\gamma~ \widehat{\mathcal O}_{\Psi}^{(1)}~.
\end{equation}
Defining the wavefunction couplings by
\begin{equation}
	\pd_{\mu_a}I_{\Psi} = \int\dd^2x~\sqrt\gamma~ \widehat{\mathcal O}_{\Psi}^{(a)}~, \qquad a=1,2~,
\end{equation}
gives
\begin{equation}\label{eq:mu lambda in r TT}
	\mu_2=-i\lambda_2~, \qquad \mu_1=\lambda_1~.
\end{equation}
Thus the relation between the real mixed-boundary-condition parameters and the wavefunction couplings is operator dependent. 
For the $\TTbar$ deformation, a real $\lambda_2$ corresponds to the imaginary wavefunction coupling $\mu_2=-i\lambda_2$, whereas on the selected root-$\TTbar$ branch $\mu_1=\lambda_1$ remains real. 
Below we work with the real mixed-boundary-condition parameters $\lambda_1$ and $\lambda_2$. 

The $\TTbar$ and root-$\TTbar$ flows generated by \eqref{eq:TTroot operators BY} commute \cite{Ebert:2023tih,Ran:2025xas}. 
We may therefore integrate them successively. 
On the branch continuously connected to $\lambda_1=\lambda_2=0$, combining the root-$\TTbar$ flow with the $\TTbar$ solution of Eqs.~\eqref{eq:solution t alpha} and \eqref{eq:solution omega alpha} gives
\begin{align}
	t_1 &= -\frac{h}{1-\lambda_2h}~, \qquad t_2 = \frac{h}{1+\lambda_2h}~, \label{eq:t solution in TTbar} \\
	\omega_1 &= e^{-2\lambda_1}(1+\lambda_2h)^2~, \qquad \omega_2 = e^{2\lambda_1}(1-\lambda_2h)^2~. \label{eq:omega solution in TTbar}
\end{align}
Together with \eqref{eq:newCJ} and \eqref{eq:newCA}, these relations determine $h$ and $\theta$ in terms of $\lambda_1$, $\lambda_2$, $R$, and the undeformed spectrum. 

Introducing
\begin{equation}
	K \equiv \sqrt{(E^{[0]})^2+(J^{[0]})^2}~,
\end{equation}
and
\begin{equation}
	A_1 \equiv K\sinh 2\lambda_1+E^{[0]}\cosh 2\lambda_1~, \qquad A_2 \equiv \pd_{\lambda_1}A_1 = 2\left( K\cosh 2\lambda_1+E^{[0]}\sinh 2\lambda_1 \right)~,
\end{equation}
the fixed-state conditions \eqref{eq:newCJ} and \eqref{eq:newCA} give
\begin{align}
	h = & \frac{ R+2A_1\lambda_2 -\sqrt{ R^2+4A_1\lambda_2R -4(J^{[0]})^2\lambda_2^2 } }{ A_2\lambda_2^2}~, \label{eq:h solution TT root} \\
	\sin 2\theta = & \frac{ J^{[0]}(1-\lambda_2^2h^2) }{ hR }~. \label{eq:theta solution TT root}
\end{align}
We choose the branch of \eqref{eq:h solution TT root} that is smooth as $\lambda_2\to0$. 
The corresponding energy follows from \eqref{eq:deformed energy},
\begin{equation}
	E^{[\lambda_1,\lambda_2]} = \frac{ \sqrt{ R^2+4A_1\lambda_2R -4(J^{[0]})^2\lambda_2^2 } -R }{ 2\lambda_2 }~. 
\end{equation}
Together with \eqref{eq:omega solution in TTbar}, Eqs.~\eqref{eq:h solution TT root} and \eqref{eq:theta solution TT root} determine the pseudo entropy through \eqref{eq:general pseudo entropy}. 
For nonzero angular momentum, it is convenient to retain this parametric form, since $\theta$ mixes the compact and noncompact Euclidean directions.

For $J^{[0]}=0$, we choose $\theta=0$ and use \eqref{eq:effective temperature general}. 
The horizon-regularity scale is related to the undeformed energy by
\begin{equation}\label{eq:beta TT root nonrotating}
	\beta
	=
	\sqrt{\frac{\pi\ell R}{4G E^{[0]}}}
	\sqrt{
		e^{-4\lambda_1}
		+\frac{4E^{[0]}\lambda_2}{R}e^{-2\lambda_1}
	}~.
\end{equation}
Equivalently, at fixed $\beta$,
\begin{align}
	\sqrt{\omega_1}
	&=
	\frac{2e^{-\lambda_1}}
	{1+
		\sqrt{1-e^{-2\lambda_1}\dfrac{\pi\ell\lambda_2}{G\beta^2}}}~,
	\label{eq:omega1 beta nonrot}
	\\
	\sqrt{\omega_2}
	&=
	\frac{2e^{\lambda_1}
		\sqrt{1-e^{-2\lambda_1}\dfrac{\pi\ell\lambda_2}{G\beta^2}}}
	{1+
		\sqrt{1-e^{-2\lambda_1}\dfrac{\pi\ell\lambda_2}{G\beta^2}}}~.
	\label{eq:omega2 beta nonrot}
\end{align}
We choose the square-root branch continuously connected to the undeformed solution. 
Here and below, $\lambda_1$ and $\lambda_2$ are the real mixed-boundary-condition parameters in the Brown--York convention; their wavefunction counterparts are given by \eqref{eq:mu lambda in r TT}.

For either sign of $\lambda_2$, the mixed boundary condition at $\mathcal I^+$ defines a real source--response relation wherever the square-root discriminant is nonnegative and the boundary source is nondegenerate. 
This statement does not rely on a finite-cutoff interpretation. 
In the pure $\TTbar$ case, the mixed boundary condition for $\lambda_2<0$ can be rewritten on shell as a Dirichlet condition at a finite radial cutoff in Fefferman--Graham gauge \eqref{eq:FG gauge}. \footnote{\label{fn:cutoff}
Substituting the Brown--York relation into the $\TTbar$ metric-flow solution gives
\begin{equation*}
	\gamma_{ab}=g^{(0)}_{ab}+\rho_c g^{(2)}_{ab}+\rho_c^2 g^{(4)}_{ab}~,\qquad\rho_c=-\frac{\lambda_2}{4\pi G\ell}~.
\end{equation*}
Hence $\lambda_2<0$ corresponds to $\rho_c>0$ \cite{Guica:2019nzm,Ran:2024vgl}.} 
For $\lambda_2>0$, no positive real cutoff representation exists, although the mixed boundary condition at $\mathcal I^+$ remains real up to the branch endpoint. 
A generic combined deformation with $\lambda_1\neq0$ likewise has no simple radial-cutoff interpretation. 
For $\lambda_2<0$, the discriminant in \eqref{eq:omega1 beta nonrot} and \eqref{eq:omega2 beta nonrot} is positive for real $\beta$.

For $\lambda_2>0$, the real mixed-boundary-condition branch terminates when the discriminant in \eqref{eq:omega1 beta nonrot} and \eqref{eq:omega2 beta nonrot} vanishes,
\begin{equation}
	\beta
	=
	\beta_H~,
	\qquad
	\beta_H
	=
	e^{-\lambda_1}
	\sqrt{\frac{\pi\ell\lambda_2}{G}}~.
\end{equation}
The discriminant inherits the square-root structure of the AdS $T\bar T$ result, while its interpretation here follows from the degeneration of the real de Sitter mixed-boundary-condition saddle \cite{Datta:2018thy,Aharony:2018bad,Chakraborty:2020swe}. 
In the present dS construction, however, $\lambda_2$ is the real mixed-boundary-condition parameter, while the corresponding wavefunction coupling is $\mu_2=-i\lambda_2$. 
We therefore interpret $\beta_H$ as a Hagedorn-like branch point of the real dS mixed-boundary-condition saddle, rather than as a thermal Hagedorn transition of a unitary theory with real $\TTbar$ coupling. 
The factor $e^{-\lambda_1}$ gives the root-$\TTbar$ rescaling of the endpoint.

At fixed $\lambda_1$, $\lambda_2>0$, and $R$, approaching $\beta_H$ gives $\lambda_2h\to1$ and $\omega_2\to0$. 
The source metric degenerates, while the response and horizon area diverge. 
The algebraic solution remains defined for $\beta>\beta_H$, but the endpoint is not a regular mixed-boundary-condition geometry. 
The full-interval image prescription established above has a narrower domain. 
Equations~\eqref{eq:effective temperature general}, \eqref{eq:omega1 beta nonrot}, and
\eqref{eq:omega2 beta nonrot} give
\begin{equation}
    U=\frac{2\pi R e^{-2\lambda_1}}
    {\sqrt{\beta^2-\beta_H^2}}~.
\end{equation}
Thus the no-conical-excess condition requires $\beta^2\geq\beta_H^2+R^2e^{-4\lambda_1}$.
At fixed couplings and $R$, this sector is left before $\beta$ reaches $\beta_H$; the image-selection analysis of App.~\ref{app:complex-geodesic-nonrotating} therefore does not extend to the endpoint.

Substituting \eqref{eq:omega1 beta nonrot} and \eqref{eq:omega2 beta nonrot} into \eqref{eq:nonrotating pseudo entropy beta}, we obtain
\begin{equation}\label{eq:0Jentropy}
	S^{[\lambda_1,\lambda_2]}(\phi_0) =
	\frac{\pi c_{\rm dS}}{6}
	-\frac{i c_{\rm dS}}{3}
	\ln\left[
	\frac{\beta e^{\lambda_1}}{2\pi\epsilon}
	\left(
	1+
	\sqrt{1-e^{-2\lambda_1}\frac{\pi\ell\lambda_2}{G\beta^2}}
	\right)
	\sin\left(
	\frac{\pi e^{-2\lambda_1}\phi_0}
	{\beta\sqrt{1-e^{-2\lambda_1}\dfrac{\pi\ell\lambda_2}{G\beta^2}}}
	\right)
	\right]~.
\end{equation}
The $\lambda_2$ dependence enters through the discriminant under the square root, while $\lambda_1$ controls the anisotropic rescaling of the two boundary directions. 
The square root in \eqref{eq:0Jentropy} is inherited from the mixed-boundary-condition solution and should be distinguished from the inverse-cosine sheet used to define the complex geodesic length. 

Replica constructions support the complex-extremal-surface prescription in undeformed, highly symmetric de Sitter settings \cite{Narayan:2023zen,Nanda:2025tid}. 
For $T\bar T$ deformations, Chen--Jiang--Yang \cite{Chen:2023eic} obtain an exact trace-flow/replica result for antipodal points on a sphere. 
Chang et al.~\cite{Chang:2024voo} use a CHM/effective-radius construction to treat arbitrary single intervals, including finite-size and finite-temperature settings at finite cutoff. 
Our calculation instead evaluates local first-order $T\bar T$ and root-$T\bar T$ insertions on the undeformed thermal replica cylinder and continues the result along the wavefunction-coupling trajectory \eqref{eq:mu lambda in r TT}.

On the non-rotating principal branch, the leading large-$c$ calculation in App.~\ref{app:replica pseudo entropy} reproduces the interval-dependent terms in the $O(\lambda_1,\lambda_2)$ expansion of \eqref{eq:0Jentropy}. 
The interval-independent terms agree upon matching the source-frame regulators. 
This check complements the finite-coupling constructions above, but does not construct the deformed replica saddle at finite $n$.
Beyond first order, \eqref{eq:0Jentropy} remains a prediction for the specified geodesic branch, rather than a replica-derived boundary observable.

	
\section{Conclusions}\label{sec:Conclusion}
	
We have formulated a semiclassical description of stress-tensor deformations in dS/CFT through mixed boundary conditions at future infinity. 
For derivative-free, similarity-invariant operators, the metric flow determines the coupled evolution of the boundary source and stress-tensor response. 
The deformation thereby modifies the source--response relation of the cosmological wavefunction while preserving the local bulk Einstein equations. 
The eigenvalue formulation provides a constructive implementation of this prescription. 

In Kerr-dS$_3$/CFT$_2$, the resulting holographic charges agree with the field-theory spectral flow. 
This agreement provides a nontrivial internal consistency check of the mixed-boundary dictionary. 
The same flowed boundary data determine a general holographic pseudo-entropy formula, nonperturbative in the deformation parameters within the chosen complex-geodesic branch. 
We obtained explicit expressions for the $\TTbar$ and root-$\TTbar$ deformations, relating their finite-coupling dependence to the geometry selected by the boundary conditions. 

The non-rotating analysis makes the geometric prescription precise in the sector without conical excess. 
The shortest admissible real auxiliary image is locally stable, and an independent inverse-cosine sheet choice fixes the complex de Sitter length. 
The selected complex length follows from the auxiliary length through the boundary asymptotic relation established in App.~\ref{app:complex-geodesic-nonrotating}. 
An analytically continued replica calculation reproduces the interval-dependent corrections at first order in both couplings and leading semiclassical order. 
The interval-independent terms agree after a source-frame regulator identification, which is additional input to the local replica calculation. 
These results support the prescribed pseudo entropy at the order tested and provide finite-coupling predictions beyond it. 

Several directions remain open.
	The metric-flow construction is dimension independent only as a local large-$N$ source--response relation: for derivative-free similarity-invariant operators, the eigenvalue reduction holds pointwise for inhomogeneous data in any boundary dimension, generally in an $x$-dependent local frame.
	Our holographic test is restricted to Kerr-dS$_3$/CFT$_2$.
	For $d\geq3$, the response resides in nontrivial $g^{(d)}_{ab}$ data, the Fefferman--Graham expansion does not truncate algebraically, local gravitational modes and differential constraints appear, and even-dimensional anomalies add local terms.
	A higher-dimensional dictionary therefore requires solving the mixed gravitational boundary-value problem and selecting a regular cosmological saddle and contour. Integrating the pointwise eigenvalue flow is insufficient.
	It would also be useful to relate our mixed-boundary picture to $\TTbar+\Lambda_2$ flows, static-patch holography, and stretched-horizon descriptions \cite{Anninos:2011af,Susskind:2023hnj,Coleman:2021nor,Aguilar-Gutierrez:2024nst,Chang:2025ays,Aguilar-Gutierrez:2026ogo}.
	
	The pseudo-entropy interpretation also deserves further study.
	Our complex-geodesic results for intervals in deformed Kerr-dS$_3$ should be compared with work on timelike entanglement, pseudo entropy, $\TTbar$-deformed holographic entropy, and replica approaches to de Sitter entropy
	\cite{Donnelly:2018bef,Chen:2018eqk,Grieninger:2019zts,He:2023xnb,Chang:2024voo,Zhao:2025zgm,Dai:2026tmp,Arias:2019pzy,Arias:2019zug,Arenas-Henriquez:2022pyh}. 
	Such comparisons may clarify how stress-tensor deformations reorganize transition-matrix data in the putative Euclidean dual.
	
	Finally, the construction relies on large-$N$ factorization.
	At finite $N$, stress-tensor fluctuations and contact terms should correct the mixed boundary condition at future infinity, as in general multi-trace frameworks \cite{Witten:2001ua,Berkooz:2002ug,Papadimitriou:2007sj}.
	This problem should be compared with finite-$N$ approaches to stress-tensor-deformed de Sitter holography \cite{Silverstein:2024xnr} and with freelance holography, which treats general gravitational boundary conditions and their holographic data systematically \cite{Parvizi:2025shq,Parvizi:2025wsg,Sheikh-Jabbari:2025kjd}.
	One may also ask whether nonlocal stress-tensor deformations obtained by integrating out metric degrees of freedom, or evaluating them at a saddle \cite{Kawamoto:2025oko,Li:2025lpa,Xie:2026kek}, correspond to generalized mixed boundary conditions for the cosmological wavefunction.

	
	\section*{Acknowledgments}
	
	The authors thank Yidian Chen, Wenliang Li, Zhibin Li, Xiao-Li Liu, Rongxin Miao, Jie Ren, Jia-Rui Sun, Yunfei Xie and Zi-Xuan Zhao for useful discussions. 
	H. O. is supported by the Science and Technology Development Plan Project of Jilin Province of China, Grant No. 20240101326JC.
	
	\medskip
	\noindent\textbf{Use of AI-assisted technology.}
	In preparing and revising this manuscript, the authors used OpenAI
	ChatGPT as an assistive tool for language editing, manuscript
	organization, and discussion of the presentation and consistency of
	technical arguments. The authors independently carried out and verified
	all scientific derivations, calculations, literature checks,
	interpretations, and conclusions, and take full responsibility for the
	final content of the manuscript.
	
	
	\appendix
	
	
	\section{Derivation details}

	\subsection{Metric-flow as characteristics}
	\label{app:character}
	
	We derive the source--response equations of Sec.~\ref{subsec:metric-flow} by the method of characteristics.
	The starting point is the large-$N$ generating functional, equivalently the saddle-point action $S_\lambda[\gamma]$, rather than a local Lagrangian density.
	The deformation equation takes functional Hamilton--Jacobi form,
	\begin{equation}
		\partial_\lambda S_\lambda[\gamma]
		=
		H_\lambda[\gamma,P]~,
		\qquad
		P^{ab}(x)\equiv
		\frac{\delta S_\lambda}{\delta\gamma_{ab}(x)} ,
		\label{eq:app-HJ}
	\end{equation}
	where
	\begin{equation}
		H_\lambda[\gamma,P]
		=
		\int d^dx\,\sqrt{\gamma}\,
		\mathcal O\!\left(T^a{}_b(x);\lambda\right),
		\qquad
		T^{ab}(x)
		=
		-\frac{2}{\sqrt{\gamma}}P^{ab}(x).
		\label{eq:app-Hamiltonian}
	\end{equation}
	Any nonlocality of the generating functional resides in $S_\lambda[\gamma]$ itself. The characteristic flow uses only the local dependence of the deformation Hamiltonian on $\gamma_{ab}(x)$ and $P^{ab}(x)$.
	This is the functional analogue of the method applied to deformed classical Lagrangians in \cite{Hou:2022csf}.
	The remaining Charpit equations for $S_\lambda$ and the scalar momentum $p_\lambda\equiv\partial_\lambda S_\lambda$ conjugate to $\lambda$ reconstruct the generating functional along a characteristic and are unnecessary for the source--response flow.
	
	Writing \eqref{eq:app-HJ} as
	\begin{equation}
		\mathcal F
		\equiv
		\partial_\lambda S_\lambda[\gamma]
		-
		H_\lambda[\gamma,P]
		=0~,
	\end{equation}
	gives the characteristic equations on the infinite-dimensional source space:
	\begin{equation}
		\partial_\lambda\gamma_{ab}(x)
		=
		-\frac{\delta H_\lambda}{\delta P^{ab}(x)},
		\qquad
		\partial_\lambda P^{ab}(x)
		=
		\frac{\delta H_\lambda}{\delta\gamma_{ab}(x)} ,
		\label{eq:app-characteristic-functional}
	\end{equation}
	where the second derivative is taken at fixed $P^{ab}$.
	
	The first equation gives
	\begin{equation}
		\partial_\lambda\gamma_{ab}
		=
		2\frac{\partial\mathcal O}{\partial T^{ab}} .
		\label{eq:app-flow-gamma}
	\end{equation}
	For the stress-tensor flow, \eqref{eq:def of stress tensor} gives
	\begin{equation}
		P^{ab}
		=
		-\frac12\sqrt{\gamma}\,T^{ab}.
	\end{equation}
	Taking a $\lambda$ derivative and using \eqref{eq:app-characteristic-functional} gives
	\begin{equation}
		-\frac12\sqrt{\gamma}\,
		\partial_\lambda T^{ab}
		-\frac14\sqrt{\gamma}\,
		T^{ab}\gamma^{cd}\partial_\lambda\gamma_{cd}
		=
		\frac{\delta H_\lambda}{\delta\gamma_{ab}} .
		\label{eq:app-stress-pre}
	\end{equation}
	At fixed $ P^{ab}$, the metric variation of
	$T^{cd}=-2 P^{cd}/\sqrt{\gamma}$ is
	\begin{equation}
		\left.
		\frac{\delta T^{cd}}{\delta\gamma_{ab}}
		\right|_{ P}
		=
		-\frac12 T^{cd}\gamma^{ab}.
	\end{equation}
	Therefore
	\begin{equation}
		\frac{1}{\sqrt{\gamma}}
		\frac{\delta H_\lambda}{\delta\gamma_{ab}}
		=
		\frac12\mathcal O\,\gamma^{ab}
		+ 
		\frac{\partial\mathcal O}{\partial\gamma_{ab}}
		-
		\frac12 T^{cd}\gamma^{ab}
		\frac{\partial\mathcal O}{\partial T^{cd}} .
		\label{eq:app-H-var}
	\end{equation}
	Substituting \eqref{eq:app-flow-gamma} and \eqref{eq:app-H-var} into
	\eqref{eq:app-stress-pre}, one obtains
	\begin{equation}
		\partial_\lambda T^{ab}
		=
		\left(
		T^{cd}\gamma^{ab}
		-
		T^{ab}\gamma^{cd}
		\right)
		\frac{\partial\mathcal O}{\partial T^{cd}}
		-
		2\frac{\partial\mathcal O}{\partial\gamma_{ab}}
		-
		\mathcal O\,\gamma^{ab}.
		\label{eq:app-flow-T}
	\end{equation}
	Equations \eqref{eq:app-flow-gamma} and \eqref{eq:app-flow-T} are the
	metric-flow equations quoted in the main text.
	
	For completeness, the on-shell functional can be reconstructed along the characteristic. Since
	\begin{equation}
		\partial_\lambda S_\lambda[\gamma]
		=
		H_\lambda[\gamma,P],
	\end{equation}
	one has along a characteristic
	\begin{equation}
		\frac{dS_\lambda}{d\lambda}
		=
		H_\lambda[\gamma,P]
		+
		\int d^d x\,
		P^{ab}\partial_\lambda\gamma_{ab}.
	\end{equation}
	If the characteristic connects $\gamma^{[0]}_{ab}$ at $\lambda=0$ to $\gamma^{[\lambda]}_{ab}$, then
	\begin{equation}
		S_\lambda[\gamma^{[\lambda]}]
		=
		S_0[\gamma^{[0]}]
		+
		\int_0^\lambda d\lambda'\,
		\left[
		H_{\lambda'}[\gamma(\lambda'),P(\lambda')]
		+
		\int d^d x\,
		P^{ab}(\lambda')\,
		\partial_{\lambda'}\gamma_{ab}(\lambda')
		\right].
	\end{equation}
	Although the main text uses only the induced source--response flow, this formula explicitly relates the deformed and undeformed saddle functionals along the characteristic.
	
	
	\subsection{Derivation for Eq.~\eqref{eq:t t0 omega}}\label{app:character II}
	
	Equations~\eqref{eq:eigenvalue flow omega} and \eqref{eq:eigenvalue flow t} imply that $\sqrt{\gamma}(t_\alpha-t_\beta)$ is flow invariant:
	\begin{equation}
		\frac{\dd}{\dd\lambda}\sqrt{\gamma}(t_{\alpha}-t_\beta)= \frac{1}{2} \sqrt{\gamma}(t_\alpha-t_\beta)\sum^{d}_{\sigma=1}\omega_\sigma^{-1}\frac{\dd\omega_\sigma}{\dd\lambda}+\sqrt{\gamma}\left(\frac{\dd t_\alpha}{\dd\lambda}-\frac{\dd t_\beta}{\dd\lambda}\right)=0~,
	\end{equation}
	Thus
\begin{equation}\label{eq:sqrt gamma t-t constant}
	\sqrt{\gamma}(t_{\alpha}-t_\beta)=\sqrt{\gamma_0}(t_{\alpha,0}-t_{\beta,0})~.
\end{equation}
At fixed initial metric, differentiating the eigenvalue flow \eqref{eq:eigenvalue flow t} with respect to $t_{\beta,0}$ and using \eqref{eq:eigenvalue flow omega} and \eqref{eq:sqrt gamma t-t constant} gives
\begin{align}
	\frac{\pd}{\pd\lambda}\left(\sqrt{\gamma}\frac{\pd t_\alpha}{\pd t_{\beta,0}}\right)=& \sqrt{\gamma}\sum_{\sigma=1}^{d}\frac{\pd\mathcal{O}}{\pd t_\sigma}\frac{\pd t_\alpha}{\pd t_{\beta,0}}+\sqrt{\gamma}\frac{\pd}{\pd t_{\beta,0}}\left(-\mathcal{O}+\sum^{d}_{\sigma=1}(t_\sigma-t_\alpha)\frac{\pd\mathcal{O}}{\pd t_\sigma}\right)\notag \\
	=& \sqrt{\gamma}\sum^{d}_{\sigma=1}(t_\sigma-t_\alpha)\frac{\pd}{\pd t_{\beta,0}} \frac{\pd\mathcal{O}}{\pd t_\sigma} \notag\\
	=& \frac{1}{2} \sqrt{\gamma} \sum^{d}_{\sigma=1}(t_\sigma-t_\alpha)\frac{\pd}{\pd t_{\beta,0}} \frac{\pd}{\pd\lambda}\log\omega_\sigma \notag\\
	=& \frac{1}{2} \frac{\pd}{\pd\lambda} \left[\sqrt{\gamma_0}\sum^{d}_{\sigma=1}(t_{\sigma,0}-t_{\alpha,0})\frac{\pd}{\pd t_{\beta,0}} \log\omega_\sigma\right]~.
\end{align}
Integrating from $\lambda=0$, with $\omega_\sigma(0)=1$ and $t_\alpha(0)=t_{\alpha,0}$, yields
\begin{equation}
	\sqrt{\gamma}\frac{\pd t_\alpha}{\pd t_{\beta,0}}= \sqrt{\gamma_0}~\delta^\alpha_\beta+\frac{1}{2} \sqrt{\gamma_0} \sum^{d}_{\sigma=1} (t_{\sigma,0}-t_{\alpha,0}) \frac{\pd}{\pd t_{\beta,0}} \log \omega_\sigma~,
\end{equation}
which completes the proof of Eq.~\eqref{eq:t t0 omega}.
	
	
\section{Temperature for Kerr-dS}\label{app:Temperature}
	
We derive the temperature of \eqref{eq:deformed dS3 bulk geometry} on the branch $\mathcal N>0$, $\omega_1>0$, and $\cos\theta>0$.
	
	Define
	\begin{align}
		A(\rho) =& \frac{\omega_2 \cos^2\theta (1-\rho \mathcal{N}^2)^2
			+ \omega_1 \sin^2\theta  (1+\rho \mathcal{N}^2)^2}
		{\rho \omega_1 \omega_2}~, \label{eq:def A app}\\
		B(\rho) =& \frac{\omega_2 \sin^2\theta (1-\rho \mathcal{N}^2)^2
			+ \omega_1 \cos^2\theta (1+\rho \mathcal{N}^2)^2}
		{\rho \omega_1 \omega_2}~, \\
		C(\rho) =& \frac{\sin 2\theta \left[
			\omega_2 (1-\rho \mathcal{N}^2)^2 - \omega_1 (1+\rho \mathcal{N}^2)^2
			\right]}
		{\rho \omega_1 \omega_2}~,\label{eq:def C app}
	\end{align}
	Then \eqref{eq:deformed dS3 bulk geometry} becomes
	\begin{equation}\label{eq:dS in ABC app}
		\dd s^2=-\frac{\ell^2}{4\rho^2}\dd\rho^2 + A(\rho) \dd T^2+B(\rho) \dd\phi^2+C(\rho) \dd T \dd \phi~,
	\end{equation}
	where $\mathcal{N}^2=\sqrt{\mathcal{L}\bar{\mathcal L}}$ as in \eqref{eq:def for cN}, and the horizon lies at
	\begin{equation}\label{eq:def rhoh app}
		\rho_h\equiv \frac{1}{\mathcal{N}^2}~.
	\end{equation}
	
	Introduce the co-rotating coordinate
	\begin{equation}
		\tilde{\phi}=\phi - \tan\theta~T~, 
	\end{equation}
	in which \eqref{eq:dS in ABC app} becomes
	\begin{align}
		\dd s^2= & -\frac{\ell^2}{4\rho^2}\dd\rho^2+\Big(A(\rho)+\tan\theta C(\rho)+\tan^2\theta B(\rho)\Big)\dd T^2 \notag \\
		&+B(\rho) \dd\tilde{\phi}^2+\Big(C(\rho)+2\tan\theta B(\rho)\Big)\dd T\dd\tilde{\phi}~.
	\end{align}
At the horizon, one obtains 
\begin{equation}
	g_{T\tilde{\phi}}|_{\rho=\rho_h}=0~,
\end{equation}
At fixed $\tilde\phi$, $\partial_T$ corresponds to the horizon generator $\chi=\partial_T+\tan\theta\,\partial_\phi$ in the original coordinates. 
Its thermal period follows from local regularity of the analytically continued $(\rho,T)$ plane,
\begin{equation}\label{eq:effective 2d Min app}
	\dd \tilde{s}^2=-\frac{\ell^2}{4\rho^2}\dd\rho^2+ \Big(A(\rho)+\tan\theta C(\rho)+\tan^2\theta B(\rho)\Big)\dd T^2~.
\end{equation}
	
	Near the horizon, set
	\begin{equation}
		\delta_h=1-\frac{\rho}{\rho_h}
	\end{equation}
	with infinitesimal $\delta_h$, so $\dd\rho=-\rho_h\dd\delta_h$.
	Equations~\eqref{eq:def A app}--\eqref{eq:def C app} and \eqref{eq:def rhoh app} then reduce \eqref{eq:effective 2d Min app} to
	\begin{equation}\label{eq:Min near horizon}
		\dd \tilde{s}^2 \approx -\frac{\ell^2}{4}\dd\delta_h^2+\frac{\mathcal{N}^2 \delta_h^2}{\omega_1\cos^2\theta} \dd T^2~.
	\end{equation}
Here $\approx$ denotes the leading near-horizon expansion.
Continuing $T$ at fixed $\tilde\phi$,
\begin{equation}
    T_E\equiv iT~,
\end{equation}
gives a negative-definite two-dimensional metric.
Its overall sign does not affect the conical regularity condition: 
\begin{equation}
	-\dd \tilde{s}^2_{E}\approx \frac{\ell^2}{4}\dd\delta_h^2+\frac{\mathcal{N}^2 \delta_h^2}{\omega_1\cos^2\theta}\dd T^2_{E}~.
\end{equation}
In terms of the proper radial coordinate $r_{\rm prop}$,
	\begin{equation}
		-\dd\tilde{s}^2_{E}\approx \dd r_{\rm prop}^2+\frac{4\mathcal{N}^2}{\ell^2\omega_1\cos^2\theta}r_{\rm prop}^2 \dd T^2_{E}~,
	\end{equation}
	where
	\begin{equation}
		r_{\rm prop}=\frac{\ell}{2}\delta_h~.
	\end{equation}
	Smoothness fixes the Euclidean-time period $\beta$ through
	\begin{equation}
		\frac{2\mathcal{N}}{\ell\sqrt{\omega_1}\cos\theta}\beta=2\pi~,
	\end{equation}
and hence the Kerr-dS temperature is
\begin{equation}\label{eq:TdS in app}
	\mathscr T_{\rm dS}=\frac{1}{\beta}=\frac{\mathcal{N}}{\pi\ell\sqrt{\omega_1}\cos\theta}~.
\end{equation}
In the non-rotating sector, $\theta=0$ and
\begin{equation}
	\beta=\frac{\pi\ell\sqrt{\omega_1}}{\mathcal{N}}~.
\end{equation}
This completes the derivation of Eq.~\eqref{eq:effective temperature general}.

	
	\section{Complex geodesics and saddle selection in the non-rotating sector}
	\label{app:complex-geodesic-nonrotating}
	
	We derive the complex geodesics governing non-rotating pseudo entropy and analyze their saddle structure.
	Set $J^{[0]}=0$ and choose
	\begin{equation}
		\theta=0~,
	\end{equation}
	on the regular mixed-boundary-condition branch $\omega_2>0$.
	For short intervals in the non-rotating conical-defect sector, the principal image is a unique smooth geodesic at fixed endpoints.
	It has positive second variation and minimal auxiliary Euclidean length.
	After selecting the real auxiliary image, we separately choose the principal inverse-cosine sheet for its complex de Sitter length.
	This gives a definite saddle within the family represented by real auxiliary geodesics.
	
	\subsection{Auxiliary hyperbolic cone and image saddles}
	
	For $\theta=0$, an interval at fixed $T$ is also at fixed $\tau$ in the diagonal frame \eqref{eq:tau x to T phi}, with
	\begin{equation}
		\Delta\tau=0~,
		\qquad
		\Delta x=\frac{\phi_0}{\sqrt{\omega_2}}~.
	\end{equation}
	The relevant two-dimensional bulk metric is therefore the $\tau=\text{constant}$ slice
	\begin{equation}
		\dd s^2
		=
		-\frac{\ell^2}{4\rho^2}\dd\rho^2
		+
		\frac{(1+\rho\mathcal N^2)^2}{\rho}\dd x^2~,
		\label{eq:app-nonrotating-slice-rho}
	\end{equation}
	where
	\begin{equation}
		\mathcal N\equiv(\mathcal L\bar{\mathcal L})^{1/4}
		=\frac12\sqrt{r_-^2+r_+^2}>0~.
	\end{equation}
	No real spacelike geodesic connects distinct points on $\mathcal I^+$ along the real $\rho$ contour.
	To complexify the saddle, introduce
	\begin{equation}
		y=\frac{1+\rho\mathcal N^2}{i\sqrt\rho}~.
	\end{equation}
	Then \eqref{eq:app-nonrotating-slice-rho} becomes
	\begin{equation}
		\dd s^2
		=-
		\left(
		\frac{\ell^2\dd y^2}{y^2+4\mathcal N^2}
		+y^2\dd x^2
		\right)~.
	\end{equation}
	Thus a real geodesic of the auxiliary metric represents the complex de Sitter geodesic:
	\begin{equation}
		\dd s_E^2
		=
		\frac{\ell^2\dd y^2}{y^2+4\mathcal N^2}
		+y^2\dd x^2~.
		\label{eq:app-auxiliary-H2}
	\end{equation}
	Defining
	\begin{equation}
		y=2\mathcal N\sinh\sigma~,
		\qquad
		u=\frac{2\mathcal N}{\ell}x~,
	\end{equation}
	we obtain
	\begin{equation}
		\dd s_E^2
		=
		\ell^2\left(
		\dd\sigma^2+\sinh^2\!\sigma\,\dd u^2
		\right)~.
		\label{eq:app-hyperbolic-polar}
	\end{equation}
	Locally, this is the hyperbolic plane with Gaussian curvature $K_E=-1/\ell^2$.
	Globally, $\phi\sim\phi+R$ implies
	\begin{equation}
		u\sim u+U~,
		\qquad
		U\equiv
		\frac{2\mathcal N R}{\ell\sqrt{\omega_2}}~.
		\label{eq:app-cone-angle}
	\end{equation}
	Thus the auxiliary geometry is a hyperbolic cone of opening angle $U$, smooth globally only when $U=2\pi$.
	In the undeformed non-rotating geometry, $\mathcal N=r_+/2$ and $U=r_+R/\ell$. The value $U=2\pi$ gives smooth pure dS$_3$, while Kerr-dS$_3$ conical defects have $0<U<2\pi$
	\cite{Balasubramanian:2001nb}.
	
	Fix a lift of the first endpoint at $u=0$. The second endpoint's lifts are separated from it by
	\begin{equation}
		\Delta u_n
		=u_0+nU
		=\frac{2\mathcal N}{\ell\sqrt{\omega_2}}
		(\phi_0+nR)~,
		\qquad
		n\in\mathbb Z~,
		\label{eq:app-image-separation}
	\end{equation}
	where
	\begin{equation}
		u_0
		=\frac{2\mathcal N\phi_0}
		{\ell\sqrt{\omega_2}}~,
		\qquad
		0<u_0<U
	\end{equation}
	for $0<\phi_0<R$.  We denote the absolute lifted separation by
	\begin{equation}
		\delta_n\equiv|\Delta u_n|~.
	\end{equation}
For each image satisfying
\begin{equation}\label{eq:app-admissible-image}
	0<\delta_n<\pi~,
\end{equation}
there is a unique smooth real auxiliary geodesic avoiding the cone apex.
For endpoints on the asymptotic boundary, centering the geodesic at $u=0$ gives
\begin{equation}\label{eq:app-geodesic-profile-sigma}
	\tanh\sigma\cos u=\cos\frac{\delta_n}{2}~.
\end{equation}
Equivalently,
	\begin{equation}
		y_n^2(u)
		=
		\frac{
			4\mathcal N^2\cos^2(\delta_n/2)
		}{
			\cos^2u-\cos^2(\delta_n/2)
		}~,
		\label{eq:app-geodesic-profile-y}
	\end{equation}
	with turning point
	\begin{equation}
		y_{\ast,n}
		=2\mathcal N\cot\frac{\delta_n}{2}>0~.
		\label{eq:app-turning-point-image}
	\end{equation}
	At $\delta_n=\pi$, $y_{\ast,n}=0$ and the geodesic reaches the cone apex.
	For $U=2\pi$, the apex is regular and this is an ordinary diameter of $H^2$.
	For $U<2\pi$, it is a conical singularity, so an apex-crossing curve requires an additional defect or homology prescription.
	No smooth real geodesic in this lifted class avoids the apex for $\delta_n>\pi$.
	
For endpoints at $\sigma=\sigma_c$ with angular separation $\delta_n$, the exact auxiliary length is 
\begin{align}
	\cosh\frac{D_{E,n}}{\ell}&={}\cosh^2\sigma_c-\sinh^2\!\sigma_c\cos\delta_n\notag\\
		&={}1+2\sinh^2\!\sigma_c\sin^2\frac{\delta_n}{2}~.\label{eq:app-exact-image-length}
\end{align}
	With $y_c=2\mathcal N\sinh\sigma_c\simeq\ell/\epsilon$ near the boundary, the regulated asymptotic length is
	\begin{equation}
		D_{E,n}^{\rm reg}
		=
		2\ell\log\left[
		\frac{\ell}{\mathcal N\epsilon}
		\sin\frac{\delta_n}{2}
		\right]~,
		\qquad
		0<\delta_n<\pi~.
		\label{eq:app-DE-reg}
	\end{equation}
	For $n=0$, this reduces to the auxiliary length used in
	Eq.~\eqref{eq:nonrotating pseudo entropy}.
	
	\subsection{Local stability and competition among images}
	
	We analyze stability at finite cutoff before taking the boundary limit.
	Let $\gamma_n$ be a smooth lifted geodesic parametrized by auxiliary arclength $s$, and consider a fixed-endpoint normal variation
	\begin{equation}
		\delta X^\mu=f(s)n^\mu~,
		\qquad
		f\big|_{\partial\gamma_n}=0~.
	\end{equation}
	The index form for the length functional is
	\begin{equation}
		\delta^2D_{E,n}
		=
		\int_{\gamma_n}\dd s\,
		\left[
		\left(\frac{\dd f}{\dd s}\right)^2
		-K_E f^2
		\right]
		=
		\int_{\gamma_n}\dd s\,
		\left[
		\left(\frac{\dd f}{\dd s}\right)^2
		+\frac{f^2}{\ell^2}
		\right]>0
		\label{eq:app-second-variation}
	\end{equation}
	for every nonzero normal variation. Tangential variations are pure reparametrizations.
	Thus every geodesic satisfying \eqref{eq:app-admissible-image} is nondegenerate, has no conjugate point, and strictly minimizes the real auxiliary length locally within its lifted homotopy class.
	Together with \eqref{eq:app-geodesic-profile-sigma}, this also proves uniqueness in that class.
	
	The exact length \eqref{eq:app-exact-image-length} is strictly increasing
	with the lifted separation in the admissible range:
	\begin{equation}
		\frac{\pd D_{E,n}}{\pd\delta_n}
		=
		\ell\,
		\frac{\sinh^2\!\sigma_c\sin\delta_n}
		{\sinh(D_{E,n}/\ell)}
		>0~,
		\qquad
		0<\delta_n<\pi~.
		\label{eq:app-length-monotonicity}
	\end{equation}
	Equivalently, in the boundary limit,
	\begin{equation}
		\frac{\pd D_{E,n}^{\rm reg}}{\pd\delta_n}
		=\ell\cot\frac{\delta_n}{2}>0~.
	\end{equation}
	This monotonicity selects the principal image for short intervals. For
	\begin{equation}
		0<\phi_0<\frac{R}{2}~,
		\label{eq:app-short-physical-interval}
	\end{equation}
	one has
	\begin{equation}
		|\phi_0|<|\phi_0+nR|~,
		\qquad n\neq0~,
	\end{equation}
	and consequently
	\begin{equation}
		\delta_0<\delta_n~,
		\qquad n\neq0~.
	\end{equation}
	Provided $\delta_0<\pi$, the $n=0$ geodesic is the unique shortest smooth real auxiliary image.
	In the conical-defect range $0<U\leq2\pi$, \eqref{eq:app-short-physical-interval} implies $\delta_0=u_0<U/2\leq\pi$, so the principal image is automatically admissible.
	
	For $0<\phi_0<R$, the nearest lift is
	\begin{equation}
		n_{\rm near}
		=
		\begin{cases}
			0~, & 0<\phi_0<R/2~,\\[2mm]
			-1~, & R/2<\phi_0<R~.
		\end{cases}
		\label{eq:app-nearest-image}
	\end{equation}
	At $\phi_0=R/2$, the $n=0$ and $n=-1$ lifts have equal separation $\delta_0=\delta_{-1}=U/2$.
	For $U<2\pi$, they are distinct equal-length geodesics passing opposite sides of the defect. For $U=2\pi$, they are the same unoriented diameter of the smooth disk.
	Thus the direct/complementary exchange occurs at
	\begin{equation}
		\phi_0=\frac{R}{2}~,
	\end{equation}
	whereas $\delta_n=\pi$ separately marks contact with the cone apex.
	The conditions coincide only for the smooth case $U=2\pi$.
	
	Equation~\eqref{eq:app-nearest-image} selects the shortest \emph{connected auxiliary image} in the conical-defect range.
	For large intervals, a complete pseudo-entropy phase diagram may also depend on replica homology, horizon or defect components, and the gravitational integration cycle.
	The image competition above is therefore only a controlled connected-saddle analysis and does not by itself establish a general large-interval transition.
	
	\subsection{Selected inverse-cosine sheet and the complex de Sitter length}
	
For endpoints at the physical cutoff $\rho=\rho_0>0$, define $q=\rho_0\mathcal N^2$.
In the non-rotating sector,
\eqref{eq:cos geodesic deformed} gives
\begin{equation}\label{eq:app-cos-complex-length}
    Z_n\equiv\cos\frac{L_{{\rm dS},n}}{\ell} =1-\frac{(1+q)^2}{2q}\sin^2\frac{\delta_n}{2}~.
\end{equation}
For $Z_n<-1$, the inverse-cosine branches are
\begin{equation}\label{eq:app-all-arccos-branches}
    L_{{\rm dS},n;k,\pm}=(2k+1)\pi\ell \pm i\ell\,\operatorname{arccosh}(-Z_n)~, \qquad k\in\mathbb Z~.
\end{equation}
We select the principal inverse-cosine sheet,
\begin{equation}\label{eq:app-principal-sheet}
    L_{{\rm dS},n} \equiv L_{{\rm dS},n;0,-} =\pi\ell-i\ell\,\operatorname{arccosh}(-Z_n)~.
\end{equation}
This sheet choice is additional to the selection of the real auxiliary image.
The plus sign gives the complex-conjugate branch, while $k$ labels inverse-cosine monodromies.

The auxiliary length in \eqref{eq:app-exact-image-length} is
\begin{equation}
    D_{E,n}
    =2\ell\,\operatorname{arsinh}\left[
        \frac{y_c}{2\mathcal N}\sin\frac{\delta_n}{2}
    \right].
\end{equation}
It is not equal to
$\ell\,\operatorname{arccosh}(-Z_n)$ at finite cutoff.
With asymptotically matched cutoffs
$y_c\sim\rho_0^{-1/2}=\ell/\epsilon$, however,
\begin{equation}
    L_{{\rm dS},n} =\pi\ell-iD_{E,n}+O(\rho_0), \qquad \rho_0\to0,
\end{equation}
at fixed $0<\delta_n<\pi$.

This construction does not minimize the real part, imaginary part, or modulus of $L_{\rm dS}$. 
It first orders the real auxiliary representatives, whose lengths are real and whose second variations can be tested, and only then evaluates the corresponding complex length on the chosen sheet. 
Determining whether intrinsically complex saddles have nonzero intersection numbers requires a full gravitational thimble analysis. Our result concerns only the component connected to the real auxiliary problem. 
	
For the principal short-interval image $n=0$, the regulated boundary asymptotics of \eqref{eq:app-principal-sheet}, with terms vanishing as $\epsilon\to0$ omitted, are
\begin{equation}\label{eq:app-complex-length}
	L_{\rm dS}^{\rm reg}=\pi\ell-2i\ell\log\left[\frac{\ell}{\mathcal N\epsilon}\sin\left(\frac{\mathcal N\phi_0}{\ell\sqrt{\omega_2}}\right)\right]~.
\end{equation}
For $0<u_0<\pi$, the logarithm has positive real argument and no additional monodromy. 
Thus
\begin{equation}\label{eq:app-pseudo-entropy-nonrotating}
	S_A^{[\lambda]}=\frac{L_{\rm dS}^{\rm reg}}{4G}=\frac{\pi c_{\rm dS}}{6}-\frac{ic_{\rm dS}}{3}\log\left[\frac{\ell}{\mathcal N\epsilon}\sin\left(\frac{\mathcal N\phi_0}{\ell\sqrt{\omega_2}}\right)\right]~,
\end{equation}
where
\begin{equation}
	c_{\rm dS}=\frac{3\ell}{2G}~.
\end{equation}
This reproduces \eqref{eq:nonrotating pseudo entropy}. 
The principal inverse-cosine sheet \eqref{eq:app-principal-sheet}, not minimization of a complex length, fixes the real term $\pi c_{\rm dS}/6$.
	
\subsection{A real auxiliary representative and stability under the flow}
	
	Continuing the Fefferman--Graham coordinate through the asymptotic boundary gives a useful representation of the selected saddle.
	The physical de Sitter region has $\rho>0$, whereas its auxiliary real representative lies on
	\begin{equation}
		\rho=-\zeta~,
		\qquad
		\zeta>0~.
	\end{equation}
	The constant-$\tau$ slice becomes
	\begin{equation}
		\dd s^2
		=-
		\left[
		\frac{\ell^2}{4\zeta^2}\dd\zeta^2
		+
		\frac{(1-\zeta\mathcal N^2)^2}{\zeta}\dd x^2
		\right]~.
		\label{eq:app-negative-rho-metric}
	\end{equation}
	For a lifted separation
	\begin{equation}
		\Delta x_n
		=\frac{\phi_0+nR}{\sqrt{\omega_2}}~,
	\end{equation}
	and after centering the endpoints at
	$x=\pm|\Delta x_n|/2$, the corresponding curve is
	\begin{equation}
		\zeta_n(x)
		=
		\frac{
			\cos\left(\dfrac{2\mathcal N x}{\ell}\right)
			-
			\cos\left(\dfrac{\mathcal N|\Delta x_n|}{\ell}\right)
		}{
			\mathcal N^2\left[
			\cos\left(\dfrac{2\mathcal N x}{\ell}\right)
			+
			\cos\left(\dfrac{\mathcal N|\Delta x_n|}{\ell}\right)
			\right]
		}~.
		\label{eq:app-negative-rho-geodesic}
	\end{equation}
	Thus the complex de Sitter saddle is represented by a real geodesic of the positive-definite metric inside the brackets of \eqref{eq:app-negative-rho-metric}, followed by continuation of its length.
	It is real on the continued $\rho<0$ sheet but not a real spacelike curve in the original Lorentzian geometry.
	More generally, contours deformable into one another in the complex time plane without crossing a singularity represent the same analytic saddle, not independent competitors \cite{Narayan:2026wzp}.
	
	The global auxiliary cone data simplify along the fixed-state flow of Sec.~\ref{sec:match spectrum}.
	In the non-rotating sector, Eqs.~\eqref{eq:def for cN} and \eqref{eq:A constraint} imply
	\begin{equation}
		U
		=
		\frac{2\mathcal N R}{\ell\sqrt{\omega_2}}
		=
		2\sqrt{\frac{4\pi G}{\ell}}\,
		\frac{R\sqrt h}{\sqrt{\omega_2}}
		=
		2\sqrt{\frac{4\pi G}{\ell}}\,C_A~.
		\label{eq:app-cone-angle-CA}
	\end{equation}
	Since $C_A$ is held fixed,
	\begin{equation}
		\dd U=0
		\label{eq:app-cone-angle-invariant}
	\end{equation}
	along the fixed-state trajectory.
	The deformation changes the source-frame normalization of regulated lengths but not the cone opening angle or the topological classification of image saddles.
	At fixed $\phi_0/R$, the direct/complementary exchange therefore remains at $\phi_0/R=1/2$.
	This invariance of $U$ is specific to the fixed-state flow.
	Along a canonical trajectory where the thermal state varies at fixed $\beta$, both the opening angle and the admissible smooth images may change, although the two nearest lifts remain equal at $\phi_0/R=1/2$.
	
Within the non-rotating no-conical-excess sector, the auxiliary analysis identifies the shortest admissible connected image and establishes its local stability. 
We then choose the principal inverse-cosine sheet to define its complex de Sitter length. 
Dominance in the wavefunction replica functional remains undetermined, including competition with disconnected or intrinsically complex saddles. 
	
	\section{Perturbative replica check of the deformed pseudo entropy}
	\label{app:replica pseudo entropy}
	
	We test the first-order expansion of \eqref{eq:0Jentropy} by an analytically continued replica calculation.
	The $O(\lambda_2)$ and $O(\lambda_1)$ corrections are computed separately on the non-rotating principal branch at leading semiclassical order.
	Starting from the undeformed Euclidean CFT on an $n$-sheeted thermal cylinder, we continue the result to de Sitter pseudo entropy.
	This calculation provides a perturbative boundary check of the complex-geodesic prescription.
	Mixed $O(\lambda_1\lambda_2)$ terms, rotation, and subleading quantum corrections lie beyond our scope.
	
	\subsection{Replica functional and scope}
	
	Let $\widetilde\rho_A$ be the unnormalized reduced operator prepared by a Euclidean path integral:
	\begin{equation}
		\label{eq:CFT partition app}
		Z_1=\tr\widetilde\rho_A~,
		\qquad
		\rho_A=\frac{\widetilde\rho_A}{Z_1}~,
		\qquad
		Z_n=\tr\widetilde\rho_A^{\,n}~.
	\end{equation}
	The normalized R\'enyi traces and entropy are
	\begin{equation}
		\label{eq:entanglement renyi app}
		\tr\rho_A^n=\frac{Z_n}{Z_1^n}~,
		\qquad
		S_A=-\left.\pd_n\log\frac{Z_n}{Z_1^n}\right|_{n=1}~,
	\end{equation}
	where $Z_n$ is the partition function on the branched manifold $\mathcal M_n(A)$ obtained by cyclically gluing $n$ copies along $A$
	\cite{Calabrese:2004eu}.
	
	For states $|\psi\rangle$ and $|\chi\rangle$, pseudo entropy is instead defined from the normalized transition operator
	\begin{equation}
		\mathcal T^{\psi|\chi}
		\equiv
		\frac{|\psi\rangle\langle\chi|}{\langle\chi|\psi\rangle}~,
		\qquad
		\mathcal T_A^{\psi|\chi}
		\equiv
		\tr_{\bar A}\mathcal T^{\psi|\chi}~,
	\end{equation}
	through \cite{Nakata:2020luh}
	\begin{equation}
		S^{\rm pseudo}_{A,n}
		=\frac{1}{1-n}\log\tr\!\left[
		\left(\mathcal T_A^{\psi|\chi}\right)^n\right]~,
		\qquad
		S_A^{\rm pseudo}
		=-\tr\!\left(\mathcal T_A^{\psi|\chi}
		\log\mathcal T_A^{\psi|\chi}\right).
		\label{eq:pseudo entropy transition app}
	\end{equation}
	 In dS/CFT the natural semiclassical boundary
	functional is the cosmological wavefunction $\Psi_{\rm dS}$.  When its
	preparation contour can be cut into two pieces and $n$ copies are cyclically
	sewn along $A$, the corresponding wavefunction replica ratio is formally
	\begin{equation}
		\tr\!\left[
		\left(\mathcal T_A^{\psi|\chi}\right)^n\right]
		=
		\frac{\Psi_{{\rm dS},n}[\mathcal M_n(A)]}
		{\Psi_{{\rm dS},1}[\mathcal M_1]^n}~,
		\label{eq:wavefunction replica app}
	\end{equation}
	and hence
	\begin{equation}
		S_A^{\rm pseudo}
		=-\left.\pd_n\log
		\frac{\Psi_{{\rm dS},n}}
		{\Psi_{{\rm dS},1}^{\,n}}
		\right|_{n=1}
		=\left.(1-n\pd_n)\log\Psi_{{\rm dS},n}\right|_{n=1}~.
		\label{eq:wavefunction pseudo entropy app}
	\end{equation}
	Replicating a single cosmological wavefunction rather than $|\Psi_{\rm dS}|^2$ naturally yields a transition-matrix quantity \cite{Doi:2022iyj,Nanda:2025tid}.
	
	Semiclassically, $\Psi_{{\rm dS},n}\simeq e^{iI_n^{\rm dS}}$.
	Under the usual Lewkowycz--Maldacena assumptions---replica symmetry, a smooth bulk cover, and continuation away from integer $n$---the $\mathbb Z_n$ quotient contains a codimension-two fixed locus of effective tension
	\begin{equation}
		T_n=\frac{n-1}{4nG}~.
	\end{equation}
If the replica family satisfies the mixed boundary conditions and admits the chosen analytic continuation, the $n\to1$ Lewkowycz--Maldacena prescription yields
\begin{equation}\label{eq:dS LM app}
	S_A^{\rm pseudo}=\left.(1-n\pd_n)iI_n^{\rm dS}\right|_{n=1}=\frac{\operatorname{Length}(\Gamma_A)}{4G}~,
\end{equation}
where the length is generally complex in de Sitter space \cite{Lewkowycz:2013nqa,Nanda:2025tid}.
Explicit smooth de Sitter replica saddles support this relation in undeformed, highly symmetric no-boundary settings \cite{Nanda:2025tid}, but do not provide a finite-$n$ saddle for a generic interval in stress-tensor-deformed Kerr-dS$_3$.  
	
	We therefore take a more limited route: evaluate first-order insertions on an auxiliary Euclidean CFT replica geometry, then continue the $n\to1$ result to the de Sitter observable.
	
	\subsection{First-order deformation on the replicated cylinder}
	
	Consider the undeformed Euclidean CFT on the thermal cylinder
	\begin{equation}
		w=\sigma+i\tau~,
		\qquad
		\tau\sim\tau+\beta~,
		\qquad
		A=[0,L]\quad\text{at}\quad\tau=0~.
	\end{equation}
	To treat the two deformations separately, fix $a\in\{1,2\}$, set the other coupling to zero, and denote the remaining coupling by $\lambda\equiv\lambda_a$.
	Let
	\begin{equation}
		Z_{n,\lambda}=Z_\lambda[\mathcal M_n(A)]~,
		\qquad
		\mathcal R_{n,\lambda}(A)
		\equiv\frac{Z_{n,\lambda}}{Z_{1,\lambda}^{\,n}}~,
	\end{equation}
	so $S_A(\lambda)=-\pd_n\log\mathcal R_{n,\lambda}|_{n=1}$. Suppose the Euclidean action obeys
	\begin{equation}
		\left.\pd_{\lambda}I_\lambda\right|_{\lambda=0}
		=\int\dd^2w\,\mathcal V_a(w,\bar w)~,
	\end{equation}
	then differentiating the replica ratio gives
	\begin{align}
		\left.\pd_{\lambda}\log\mathcal R_{n,\lambda}\right|_{\lambda=0}
		={}&
		-\int_{\mathcal M_n(A)}\dd^2w\,
		\langle\mathcal V_a(w,\bar w)\rangle_n
		\notag\\
		&+n\int_{\mathcal M_1}\dd^2w\,
		\langle\mathcal V_a(w,\bar w)\rangle_1~.
		\label{eq:replica variation general app}
	\end{align}
	Replica symmetry turns the integral over $\mathcal M_n(A)$ into $n$ times the integral over one sheet $\Sigma_1$, yielding
	\begin{equation}
		\left.\pd_{\lambda}S_A\right|_{\lambda=0}
		=
		\int_{\Sigma_1}\dd^2w\,
		\left.\pd_n\langle\mathcal V_a(w,\bar w)\rangle_n
		\right|_{n=1}~.
		\label{eq:replica entropy variation app}
	\end{equation}
	The definition of $\mathcal V_a$ includes any counterterms required on the conical background. Defect-localized terms can affect interval-independent contributions.
	
	Equation~\eqref{eq:replica variation general app} also shows why the deformed finite-volume spectrum alone cannot determine interval pseudo entropy.
	The spectrum fixes homogeneous data on the unbranched cylinder, whereas the replica functional also requires the deformation operator's position-dependent one-point function on $\mathcal M_n(A)$, equivalently stress-tensor and twist-defect data near the branch points.
	Below, undeformed CFT Ward identities and the uniformization map supply these additional data.
	
	As in Sec.~\ref{subsec:ads-to-ds}, distinguish the real Brown--York tensor from the analytically continued wavefunction tensor. For the auxiliary calculation,
	\begin{equation}
		\label{eq:analytical continuation app}
		I_{\rm aux}
		\xrightarrow{\ \mathcal C\ }
		I_{\Psi}=-iS_{\rm ren}^{\rm dS}~,
		\qquad
		T^{ab}_{\rm aux}
		\xrightarrow{\ \mathcal C\ }
		T^{ab}_{\Psi}=-iT^{ab}_{\rm BY}~.
	\end{equation}
	Thus $T^{ab}_{\rm BY}=iT^{ab}_{\rm aux}$ on the continued branch. Below we write $T^{ab}_{\rm aux}$ simply as $T^{ab}$.
	Since $t_{\alpha,{\rm BY}}=it_{\alpha,{\rm aux}}$, the Brown--York operators
	$\mathcal O^{(2)}=t_{1,{\rm BY}}t_{2,{\rm BY}}$ and
	$\mathcal O^{(1)}=t_{2,{\rm BY}}-t_{1,{\rm BY}}$ become
	\begin{align}
		\mathcal O^{(2)}
		&=-t_{1,{\rm aux}}t_{2,{\rm aux}}
		=\frac12T_{ab}T^{ab}-\frac12(T^a{}_a)^2~,
		\\
		\mathcal O^{(1)}
		&=i(t_{2,{\rm aux}}-t_{1,{\rm aux}})
		=2i\sqrt{\frac12T_{ab}T^{ab}-\frac14(T^a{}_a)^2}~.
	\end{align}
	We choose the branch on which the square root is positive for the real auxiliary thermal CFT. The opposite branch reverses the sign of $\lambda_1$. In complex coordinates,
	\begin{align}
		T_{ww}&=\frac14(T_{\sigma\sigma}-T_{\tau\tau}-2iT_{\sigma\tau})~,
		&
		T_{\bar w\bar w}&=\frac14(T_{\sigma\sigma}-T_{\tau\tau}+2iT_{\sigma\tau})~,
		\\
		T_{w\bar w}&=\frac14(T_{\sigma\sigma}+T_{\tau\tau})~,
	\end{align}
	and therefore
	\begin{align}
		\frac12T_{ab}T^{ab}-\frac12(T^a{}_a)^2
		&=4(T_{ww}T_{\bar w\bar w}-T_{w\bar w}^2)~,
		\\
		\frac12T_{ab}T^{ab}-\frac14(T^a{}_a)^2
		&=4T_{ww}T_{\bar w\bar w}~.
	\end{align}
	At the undeformed CFT point, $T_{w\bar w}=0$ away from the branch points.
	Combining the deformation flow \eqref{eq:def of deformation diff eq} with
	\eqref{eq:analytical continuation app} gives
	\begin{equation}
		\left.\pd_{\lambda_2}I_{\rm aux}\right|_0
		=-4i\int\dd^2w\,
		(T_{ww}T_{\bar w\bar w}-T_{w\bar w}^2)~,
		\qquad
		\left.\pd_{\lambda_1}I_{\rm aux}\right|_0
		=4\int\dd^2w\,\sqrt{T_{ww}T_{\bar w\bar w}}~.
		\label{eq:CFT deformation insertions app}
	\end{equation}
	Equivalently, the standard auxiliary Euclidean couplings multiplying these operators are
	\begin{equation}
		\mu_2=-i\lambda_2~,
		\qquad
		\mu_1=\lambda_1~.
		\label{eq:coupling continuation app}
	\end{equation}
	Thus real de Sitter Brown--York $\lambda_2$ corresponds to an imaginary $\TTbar$ direction in the auxiliary Euclidean CFT.
	The result below is the continuation of the standard Euclidean $\TTbar$ replica correction, not the entropy of a unitary theory deformed by a real Euclidean $\TTbar$ coupling.
	The root-$\TTbar$ insertion remains real on the selected positive thermal branch.
	
	On the $n$-sheeted cover, the uniformization map
	\begin{equation}
		z=
		\left(
		\frac{e^{2\pi w/\beta}-1}
		{e^{2\pi w/\beta}-e^{2\pi L/\beta}}
		\right)^{1/n}
	\end{equation}
	together with the branch-point prescription of \cite{Chen:2018eqk} gives
	\begin{equation}
		\label{eq:onepoint replica}
		\langle T_{ww}(w)\rangle_n
		=\frac{c\pi}{12\beta^2}
		-\frac{c(n-1)}{24\pi}H(w)
		+O\!\left((n-1)^2\right)~,
	\end{equation}
	together with the antiholomorphic counterpart, where
	\begin{equation}
		H(w)=
		\frac{4\pi^2(1-e^{2\pi L/\beta})^2e^{4\pi w/\beta}}
		{\beta^2(e^{2\pi w/\beta}-e^{2\pi L/\beta})^2
			(e^{2\pi w/\beta}-1)^2}~.
	\end{equation}
At fixed endpoint regulator, we first take the replica derivative at $n=1$, continuing the square root in the root-$\TTbar$ insertion from its positive thermal value. 
We then integrate over the periodic coordinate $\tau$ at fixed $\sigma\ne0,L$. 
Finally, we perform the $\sigma$ integral with small neighborhoods of $0$ and $L$ excluded and remove the regulator. 
With this endpoint prescription, the integral is
\begin{equation}\label{eq:H integral}
	\int_{\Sigma_1}\dd^2w\,[H(w)+H(\bar w)]=\frac{8\pi^2L}{\beta}\coth\frac{\pi L}{\beta}~.
\end{equation}
This calculation does not fix additional defect-localized contributions. 

\subsection{The $\TTbar$ and root-$\TTbar$ corrections}
	
	\paragraph{The $\TTbar$ deformation.}
	At factorized semiclassical order, connected stress-tensor correlators are subleading, so
	\begin{equation}
		\langle T_{ww}T_{\bar w\bar w}\rangle_n
		=\langle T_{ww}\rangle_n\langle T_{\bar w\bar w}\rangle_n
		+O(c)~.
	\end{equation}
Using $T_{w\bar w}=0$ away from the branch points, Eq.~\eqref{eq:onepoint replica} gives
\begin{equation}
	\left.\pd_n\left\langle-4i(T_{ww}T_{\bar w\bar w}-T_{w\bar w}^2)\right\rangle_n\right|_{n=1}=\frac{ic^2}{72\beta^2}[H(w)+H(\bar w)]+O(c)~.
\end{equation}
Substitution into \eqref{eq:replica entropy variation app} yields
	\begin{equation}
		\label{eq:first order lambda2 CFT}
		\left.\pd_{\lambda_2}S_A^{\rm th}\right|_0
		=\frac{i\pi^2c^2L}{9\beta^3}
		\coth\frac{\pi L}{\beta}+O(c)~.
	\end{equation}
	This is the first-order thermal-cylinder result of \cite{Chen:2018eqk}, analytically continued in the coupling and expressed in our normalization.
	
	\paragraph{The root-$\TTbar$ deformation.}
	For the non-analytic square-root operator, choose the positive thermal branch and use leading large-$c$ factorization:
	\begin{align}
		\left\langle4\sqrt{T_{ww}(w)T_{\bar w\bar w}(\bar w)}\right\rangle_n
		& =4\sqrt{\langle T_{ww}\rangle_n
			\langle T_{\bar w\bar w}\rangle_n}+O(c^0)
		\notag\\
		& =\frac{c\pi}{3\beta^2}
		-\frac{c(n-1)}{12\pi}[H(w)+H(\bar w)]
		+O\!\left(c(n-1)^2\right)+O(c^0)~.
	\end{align}
	Equation \eqref{eq:replica entropy variation app} then gives
	\begin{equation}
		\label{eq:first order lambda1 CFT}
		\left.\pd_{\lambda_1}S_A^{\rm th}\right|_0
		=-\frac{2\pi cL}{3\beta}
		\coth\frac{\pi L}{\beta}+O(c^0)~.
	\end{equation}
	The $O(c^0)$ remainder depends on the quantum definition and renormalization of the root-$\TTbar$ operator, including connected stress-tensor correlators, branch choices, and contact terms at replica defects.
	Semiclassical factorization does not fix these contributions. See \cite{Hadasz:2024vxx} for related definitional issues.
	
	\subsection{Analytic continuation and comparison with holography}
	
	The undeformed thermal-cylinder entropy is
	\begin{equation}
		S_A^{\rm th}(L;\epsilon_{\rm rep})
		=\frac{c}{3}\log\left[
		\frac{\beta}{\pi\epsilon_{\rm rep}}
		\sinh\left(\frac{\pi L}{\beta}\right)
		\right]~.
	\end{equation}
	We implement the saddle continuation \eqref{eq:AdS to dS continuation} through
	\begin{equation}
		c\equiv c_{\rm aux}
		\longrightarrow
		c_{\Psi}=-ic_{\rm dS}~,
		\qquad
		L\longrightarrow i\phi_0~,
		\qquad
		x\equiv\frac{\pi\phi_0}{\beta}~.
		\label{eq:replica continuation dictionary app}
	\end{equation}
	For $0<x<\pi$, $\sin x>0$, and the principal logarithm gives
	\begin{equation}
		\log\sinh(ix)=\log(\sin x)+\frac{i\pi}{2}~.
	\end{equation}
	The undeformed answer therefore becomes
	\begin{equation}
		\label{eq:undeformed pseudo entropy}
		S_A^{[0,0]}(\phi_0;\epsilon_{\rm rep})
		=\frac{\pi c_{\rm dS}}{6}
		-\frac{ic_{\rm dS}}{3}
		\log\left[
		\frac{\beta}{\pi\epsilon_{\rm rep}}\sin x
		\right]~.
	\end{equation}
	Thus the principal continuation branch already fixes the real term $\pi c_{\rm dS}/6$.
	In the bulk it is the corresponding branch contribution to the complex geodesic. In explicit undeformed no-boundary replica saddles, an analogous term comes from the Euclidean segment \cite{Doi:2022iyj,Nanda:2025tid}.
	For a generic deformed Kerr-dS$_3$ interval, no finite-$n$ saddle presently establishes it as an independent transition-matrix normalization.
	
	Since $\coth(\pi L/\beta)\to-i\cot x$, Eq.~\eqref{eq:first order lambda2 CFT}
	gives
	\begin{align}
		\label{eq:first order lambda2 dS}
		S_{A,{\rm rep}}^{[0,\lambda_2]}
		(\phi_0;\epsilon_{\rm rep})
		={}&S_A^{[0,0]}(\phi_0;\epsilon_{\rm rep})
		-\frac{i\pi^2c_{\rm dS}^2\lambda_2\phi_0}{9\beta^3}
		\cot x
		\notag\\
		&+O(\lambda_2^2)
		+O\!\left(\frac{\lambda_2c_{\rm dS}}{\beta^2}\right)~.
	\end{align}
	Similarly, Eq.~\eqref{eq:first order lambda1 CFT} becomes
	\begin{align}
		\label{eq:first order lambda1 dS}
		S_{A,{\rm rep}}^{[\lambda_1,0]}
		(\phi_0;\epsilon_{\rm rep})
		={}&S_A^{[0,0]}(\phi_0;\epsilon_{\rm rep})
		+\frac{2ic_{\rm dS}\lambda_1}{3}x\cot x
		\notag\\
		&+O(\lambda_1^2)+O(\lambda_1c_{\rm dS}^{0})~.
	\end{align}
	Introduce the dimensionless combination
	\begin{equation}
		\kappa\equiv\frac{\pi\ell\lambda_2}{G\beta^2}~.
		\label{eq:kappa replica app}
	\end{equation}
	Using $c_{\rm dS}=3\ell/(2G)$, the two corrections combine into
	\begin{align}
		S_{A,{\rm rep}}^{[\lambda_1,\lambda_2]}
		(\phi_0;\epsilon_{\rm rep})
		={}&S_A^{[0,0]}(\phi_0;\epsilon_{\rm rep})
		+\frac{2ic_{\rm dS}\lambda_1}{3}x\cot x
		-\frac{ic_{\rm dS}}{3}\kappa\frac{x}{2}\cot x
		\notag\\
		&+O(\lambda_1^2,\lambda_1\kappa,\kappa^2)
		+O(\lambda_1c_{\rm dS}^{0},\kappa c_{\rm dS}^{0})~.
		\label{eq:combined replica first order app}
	\end{align}
	
	At fixed $\beta$, $\phi_0$, and mixed-boundary-condition regulator
	$\epsilon_{\rm MBC}$, the holographic formula \eqref{eq:0Jentropy} expands as
	\begin{align}
		\label{eq:holographic first order}
		S_{A,{\rm hol}}^{[\lambda_1,\lambda_2]}
		(\phi_0;\epsilon_{\rm MBC})
		={}&S_A^{[0,0]}(\phi_0;\epsilon_{\rm MBC})
		+\frac{2ic_{\rm dS}\lambda_1}{3}x\cot x
		-\frac{ic_{\rm dS}\lambda_1}{3}
		\notag\\
		&-\frac{ic_{\rm dS}}{3}\kappa
		\left(\frac{x}{2}\cot x-\frac14\right)
		+O(\lambda_1^2,\lambda_1\kappa,\kappa^2)~.
	\end{align}
	The interval-dependent terms agree exactly.
	The remaining difference between \eqref{eq:combined replica first order app} and \eqref{eq:holographic first order} is interval independent and can be accounted for by the source-frame regulator matching described below.
	
	Indeed, Eq.~\eqref{eq:omega1 beta nonrot} gives
	\begin{equation}
		\sqrt{\omega_1}
		=1-\lambda_1+\frac{\kappa}{4}
		+O(\lambda_1^2,\lambda_1\kappa,\kappa^2)~.
		\label{eq:omega1 first order app}
	\end{equation}
To compare the interval-independent terms, we adopt the source-frame regulator identification
\begin{equation}
    \epsilon_{\rm rep} =\sqrt{\omega_1}\,\epsilon_{\rm MBC}~. \label{eq:regulator matching app}
\end{equation}
This identification is motivated by the prefactor in \eqref{eq:nonrotating pseudo entropy beta}; it is not determined by the local replica insertion.
At the order controlled here,
\begin{align}
	S_A^{[0,0]}(\phi_0;\epsilon_{\rm rep}) ={}&S_A^{[0,0]}(\phi_0;\epsilon_{\rm MBC}) +\frac{ic_{\rm dS}}{3} \log\frac{\epsilon_{\rm rep}}{\epsilon_{\rm MBC}} \notag\\
	={}&S_A^{[0,0]}(\phi_0;\epsilon_{\rm MBC}) -\frac{ic_{\rm dS}\lambda_1}{3} +\frac{ic_{\rm dS}\kappa}{12} +O(\lambda_1^2,\lambda_1\kappa,\kappa^2)~. \label{eq:regulator entropy shift app}
\end{align}
With this identification, the shift \eqref{eq:regulator entropy shift app} accounts for the interval-independent difference and gives agreement with \eqref{eq:holographic first order}.
Since $\omega_1$ depends on the background, this matching relates the regulators used in the two calculations rather than defining a universal, state-independent cutoff renormalization.

The replica calculation therefore directly tests the interval-dependent corrections at leading large $c$ and first order in the couplings.
Agreement of the interval-independent terms additionally requires \eqref{eq:regulator matching app}.
The calculation neither follows from the spectrum alone nor constructs the deformed wavefunction at finite $n$. Higher orders, rotation, and the finite-$c$ definition of root-$T\bar T$ at replica defects remain open.
Related $\TTbar$ replica calculations in AdS and dS appear in \cite{Chen:2018eqk,Chen:2023eic,Chang:2024voo}.

	

\providecommand{\href}[2]{#2}\begingroup\raggedright\endgroup

\end{document}